\documentclass[useAMS,usenatbib]{mn2e}
\usepackage{epsf,rotating}
\usepackage{amsmath}
\usepackage{aas_macros}
\usepackage{amssymb}
\usepackage{tabularx,ragged2e,booktabs}
\usepackage[usenames,dvipsnames]{xcolor}
\usepackage{subfigure}
\usepackage[outdir=./]{epstopdf}
\DeclareMathOperator\erf{erfc}

\newcolumntype{C}[1]{>{\Centering}m{#1}}

\newcommand{\kmsmpc}{\kms\;{\rm Mpc}^{-1}}

\newcommand{\HI}{\ion{H}{i}}

\newcommand{\hkpc}{h^{-1}{\rm kpc}}
\newcommand{\hmpc}{h^{-1}{\rm Mpc}}
\newcommand{\lcdm}{$\Lambda$CDM}

\newcommand{\kms}{\;{\rm km}\,{\rm s}^{-1}}

\newcommand{\cms}{\;{\rm cm}^{-2}}
\newcommand{\cmc}{\;{\rm cm}^{-3}}

\newcommand{\msolar}{\;{\rm M}_{\odot}}

\newcommand{\gad}{{\sc Gadget-2}}

\newcommand{\ion}[2]{\hbox{#1\,{\sc #2}}}

\newcommand{\MHI}{M_{\rm HI}}
\newcommand{\NHI}{N_{\rm HI}}
\newcommand{\resgal}{M_*\geq1.45\times10^8\msolar}

\title[Environment/mergers on {\HI}]{The impact of environment and mergers on the {\HI} content of galaxies in hydrodynamic simulations}

\author[Rafieferantsoa et al.]{
\parbox[t]{\textwidth}{\vspace{-1cm}
Mika Rafieferantsoa$^{1,2}$, Romeel Dav\'e$^{1,2,3}$, 
Daniel Angl{\'e}s-Alc{\'a}zar$^4$, Neal Katz$^5$, Juna A. Kollmeier$^6$,
Benjamin D. Oppenheimer$^7$}
\\
\\$^1$ University of the Western Cape, Bellville, Cape Town 7535, South Africa
\\$^2$ South African Astronomical Observatories, Observatory, Cape Town 7925, South Africa
\\$^3$ African Institute for Mathematical Sciences, Muizenberg, Cape Town 7945, South Africa
\\$^4$ Department of Physics and Astronomy and CIERA, Northwestern University, Evanston, IL 60208, USA
\\$^5$ Astronomy Department, University of Massachusetts, Amherst, MA 01003, USA
\\$^6$ Observatories of the Carnegie Institute of Washington, Pasadena, CA 91101, USA
\\$^7$ University of Colorado, Boulder, CO 80309, USA
}

\begin{document}

\maketitle
%---------------------------------------------------------------------------------------------------------------------------
 \begin{abstract}

The instantaneous \HI~content of galaxies is thought to be
governed by recent accretion and environment. We examine these
effects within a cosmological hydrodynamic simulation that includes
a heuristic galactic outflow model that reproduces basic observed
trends of \HI~in galaxies. We show that this model reproduces
the observed \HI~mass function (HIMF) in bins of stellar mass, as
well as the \HI~richness (M$_{\rm HI}$/M$_*$) versus local galaxy
density. For satellite galaxies in massive ($\ga 10^{12}M_\odot$)
halos, the \HI~richness distribution is bimodal and the median drops
towards the largest halo masses.  The depletion timescale of \HI~entering
a massive halo is more rapid, in contrast to the specific star
formation rate which shows little variation in the attenuation rate
versus halo mass. This suggests that, up to the halo mass scales probed
here ($\la 10^{14}M_\odot$), star formation is mainly attenuated
by starvation, but \HI~is additionally removed by stripping once a
hot gaseous halo is present. In low mass halos, the \HI~richness
of satellites is independent of radius, while in very massive halos
they become gas-poor towards the centre, confirming the increasing
strength of the stripping with halo mass.  Mergers somewhat increase
the \HI~richness and its scatter about the mean relation, tracking
the metallicity in a way consistent with it arising from inflow
fluctuations, while star formation is significantly boosted relative
to \HI.

\end{abstract}
%---------------------------------------------------------------------------------------------------------------------------

%---------------------------------------------------------------------------------------------------------------------------
%---------------------------------------------------------------------------------------------------------------------------

%---------------------------------------------------------------------------------------------------------------------------
\section{Introduction}
\indent

The primary components of galaxies are stars and gas. The amount
of gas is a vital complement to that of stars as the fuel for new
star formation, as well as the repository for the nucleosynthetic
by-products of stellar evolution.  Therefore, the gas content is a
key probe of the life cycle of galaxies.  Recently, much progress
has been made in understanding the molecular gas content of galaxies
throughout cosmic time \citep[e.g.][]{Tacconi-13} via observations of carbon
monoxide (CO) lines and other dense gas tracers.
The other major gaseous mass component in galaxies
is in the form of atomic neutral hydrogen (\HI), which has been more
difficult to study to higher redshifts owing to the current sensitivities of radio
telescopes.  In the nearby Universe, the past decade has seen major
\HI\ surveys such as the \HI\ Parkes All-Sky
Survey~\citep[HIPASS;][]{Meyer-04} and the Arecibo Legacy Fast ALFA
Survey \citep[ALFALFA;][]{Giovanelli-05}, both able to probe galaxies
down to \HI\ masses of $\MHI\approx 10^7 M_\odot$ in a uniform
(albeit \HI-selected) sample.  The GALEX Arecibo SDSS Survey
\citep[GASS;][]{Catinella-10} relaxed the \HI\ selection, and instead
selected galaxies based on stellar mass.  While they did not probe as deeply in
 stellar mass (down to $M_*\approx 10^{10}$ M$_{\odot}$), the lack of
\HI\ selection allowed them to assess the biases associated with
such a selection technique.  With the upgraded Jansky Very Large
Array, the CHILES survey aims to probe \HI\ evolution in 21cm
emission out to sizeable look back times, to $z\sim
0.45$~\citep{Fernandez-13}.  Currently, there are two major radio
facilities that, as precursors to the Square Kilometre Array
(SKA\footnote{http://www.ska.ac.za}), will probe \HI\ to unprecedented
levels both nearby and out to intermediate redshifts.  The MeerKAT
array in South Africa\footnote{http://www.ska.ac.za/meerkat/} and
the Australian Square Kilometer Array Pathfinder both have major
\HI\ surveys planned, namely the
LADUMA\footnote{http://www.ast.uct.ac.za/laduma/Home.html} and
DINGO\footnote{http://askap.org/dingo} Surveys,
which will probe \HI\ 21cm emission in galaxies out to $z\sim 1$
and $z\sim0.4$, respectively.

To optimise these investments in observational resources,
it is important to place observations of \HI\ within our modern
framework for galaxy formation and evolution.  Decades ago, 
\citet{Haynes-Giovanelli-84} used 324 isolated galaxies and found
a more constrained correlation of the optical diameter 
of a stellar disk with the \HI\ mass than with the morphological type.
Even though {\HI}
does not directly foster star formation \citep{Kennicutt-Evans-12},
it has been shown that the mass of the stars and the atomic hydrogen
({\HI}) mass are highly correlated \citep{Cortese-11,Huang-12}.
Using HIPASS, \citet{Zwaan-05} measured the {\HI} mass function
down to \HI\ masses of $M_{\rm HI}\approx 10^7M_\odot$, although
at low masses the sample is likely biased towards high-$\MHI$
systems.  They also found dependencies on environment, in the sense
that dense regions tend to steepen the {\HI} mass function. An
improvement in sensitivity was done by \citet{Martin-10}, who observed
over 10,000 \HI-selected galaxies down to $\MHI\sim$10$^{6}\msolar$,
finding a larger {\HI} cosmic density and a larger value of the
characteristic mass, M$_{HI,*}$, from a Schechter function fit.

Interestingly, the galaxy {\HI} and star formation rate observations
together showed that the amount of gas in \HI\ is inadequate to
yield the observed stellar mass of a given galaxy at a sustained
star formation rate (SFR) at any observable redshift \citep{Sancisi-08}.
To explain this discrepancy, \citet{Hopkins-McClure-Griffiths-Gaensler-08}
came up with two possibilities:  first, there should be gas
replenishment at a rate slightly smaller than the requisite usage
(in order for the gas fraction to slowly drop with time), or second,
small galaxies are less able to retain their gas.  Mass loss from
stellar evolution can help the imbalance between the amount of infalling gas
and the actual SFR \citep{Leitner-Kravtsov-11}, although not
sufficiently in lower mass galaxies and not at epochs much earlier
than today.  The idea of continued replenishment is also inferred
from the evolution of the molecular hydrogen
content~\citep[e.g.][]{Tacconi-13}.  Indeed, current theoretical
models of galaxy formation invoke continual gas infall as a driver
for star formation
\citep{Keres-05,Finlator-Dave-08,Sancisi-08,Dekel-09,Dave-Finlator-Oppenheimer-12,Lilly-13}.
Hence, it appears that the gas content of galaxies, and particularly
the \HI\ content, could provide a probe of gas accretion from the
environment around galaxies.

Environment is apparently a key factor in determining the \HI\
content of galaxies.  \citet{Cortese-11} found an anti-correlation
of {\HI} richness (i.e.  \HI\ mass per unit stellar mass) relative
to stellar mass, which is valid in moderate to low density environments
but highly clustered galaxies tend to be more {\HI} poor 
\citep{Giovanelli-Haynes-85, Solanes-01}.  This is often thought to be
related to why galaxies in very dense environments also show low
specific star formation rates.  \citet{Pappalardo-12} found 
that environment acts to lower the {\HI} content of galaxies.
\citet{Hughes-13} also found that low \HI\ objects are mostly
found in clustered regions. Data from the ALFALFA $\alpha$.40 catalogue
\citep{Haynes-11}
showed that half of the optical sources were observed in denser
regions, but only less than one quarter of all the {\HI} detected sources
are located in clusters or groups, and most \HI-rich
galaxies live outside of group environments \citep{Hess-Wilcots-13}.

Mergers, on the other hand, have a significant impact 
on galaxy evolution.  Major mergers, if between gas-rich galaxies,
typically induce bursts of star formation~\citep{Mihos-Hernquist-96}
that move galaxies well off the so-called star formation main 
sequence of SFR versus $M_*$~\citep{Noeske-07}, and hence
are partly responsible for setting the scatter around the main sequence.
\citet{Finlator-Dave-08} argued that departures from the 
``equilibrium" mass-metallicity relation are driven by stochastic
fluctuations in accretion, of which mergers are the most extreme
example, and \citet{Dave-13} suggested that this is also reflected
in the \HI\ content at a given $M_*$.

Theoretical work on studying \HI\ is also progressing rapidly.
Initially, much of the work utilised semi-analytic models (SAMs)
based on prescriptively tying the \HI\ content to the halo mass and
merger history.  \citet{Obreschkow-09} used the \citet{DeLucia-Blaizot-07}
SAM applied to the Millennium simulation, and found that with
reasonable parameter choices they could broadly match observations
of the {\HI} mass functions for both early and late type galaxies.
\citet{Lagos-11} improved on this by adding a prescription to
separate \HI\ and $H_2$ in the {\sc galform} SAM \citep{Bower-06}
with several different recipes for $H_2$, finding substantial
differences between such recipes.  They also found rapid evolution
in \HI\ properties out to $z=2$.  Similar models by \citet{Popping-Somerville-Trager-14}
showed that such differences between $H_2$ recipes are most important
in small galaxies.  \citet{Lagos-14} further examined the origin
of \HI\ in big elliptical galaxies, finding that most of the neutral
gas in the elliptical galaxies is produced by radiative cooling
from their hot halos.  While the large number of free parameters
in SAMs makes a unique physical interpretation of the results
difficult, nonetheless there is clearly interesting progress being
made from SAMs, which are especially useful for making predictions
of large upcoming surveys.

Cosmological hydrodynamic simulations have also begun to make
predictions for the \HI\ content of galaxies.  \citet{Popping-09}
presented a simple self-shielding prescription for calculating the
\HI\ content of galaxies tuned to match observations of the total
cosmic \HI\ content, and showed that such a model produces roughly
the correct \HI\ mass function.  \citet{Duffy-12} improved on this
with a more sophisticated self-shielding model which they applied
to the Overwhelmingly Large Simulations \citep[OWLS;][]{Schaye-10}.
They obtained good agreement with the observed \HI\ mass function
down to $\MHI\sim 10^{9.5}\msolar$, but below this mass they
predicted an excess, though this was not very significant since
their mass resolution limit was only a factor of several below that;
nonetheless, this disagreement mimicked a similar discord in
the stellar mass function that is likely a result of their assumed
prescription for galactic outflows \citep{Dave-Oppenheimer-Finlator-11}.  \citet{Dave-13}
presented a model with an improved recipe for galactic outflows
that matched the stellar mass function quite well, and applying
another improved self-shielding prescription, they were able to also
match the \HI\ mass function and \HI\ richness as a function of
$M_*$.  This simulation, therefore, provides a plausible
model to study how the \HI\ content of galaxies is impacted by other
factors in greater detail.

In this paper, we build on the work of \citet{Dave-13} to study the
impact of environment and halo mass on the \HI\ content of galaxies
and its evolution across cosmic time, particularly focusing
on the fluctuations in \HI\ arising from environmental processes
in satellites as well as mergers.  This paper 
is structured as follows. In section \ref{simu}, we
begin by reviewing our simulations, particularly the outflow model
that is central to matching a variety of observations, as well as
our methodology to calculate the \HI\ content of galaxies (\ref{simu2})
and our methodology for tracking galaxies back in time (\ref{simu3}).
Sections \ref{hi} and \ref{evmerg} contain our results.
 We summarise our results and discuss
implications in section \ref{sum}.

%---------------------------------------------------------------------------------------------------------------------------
\section{Methods}\label{simu}

%---------------------------------------------------------------------------------------------------------------------------
\subsection{Simulations}\label{simu1}
\indent

The main simulation we use here is the same as in \citet{Dave-13}, which
we briefly review.
Using our enhanced version of \gad~\citep{Springel-05,
Oppenheimer-Dave-08}, we run a cosmological hydrodynamic simulation
with 512$^3$ gas particles and 512$^3$ dark matter particles having
masses of $4.5\times 10^6\msolar$ and $2.3\times 10^7\msolar$,
respectively, enclosed in a periodic box of 32$\hmpc$ comoving on
a side, and a comoving gravitational softening length of 1.25$\hkpc$ (Plummer equivalent).  
We chose a \lcdm\ cosmology consistent with the Wilkinson
Microwave Anisotropy Probe results in \citet{Hinshaw-09}, namely
$\Omega_m=0.28$, $\Omega_\Lambda=0.72$, $H_0=70 \kmsmpc$,
$h\equiv H_0/(100\kmsmpc) = 0.7$,
$\sigma_8=0.82$, $\Omega_b=0.046$, and $n_s=0.96$; these parameters
are not far from that favoured by the nine-year WMAP data
\citep{Hinshaw-13} and Planck \citep{-13}.  \gad\ employs
entropy-conserving smoothed particle hydrodynamics (SPH), which has
some deficiencies related to handling surface instabilities but
this makes little difference for the bulk properties of 
galaxies~(Huang et al. 2015, in prep.).
We include radiative cooling from
primordial~\citep{Katz-Weinberg-Hernquist-96} 
with additional cooling from metal lines assuming
photo-ionisation equilibrium~\citep{Wiersma-09},
and track the metallicity
in four elements (C, O, Si and Fe) based on enrichment from Type~II supernovae, Type~Ia
supernovae, and stellar mass loss from asymptotic giant branch
stars.  We will only cursorily be concerned with the metals in this
paper, and hence we refer the reader to \citet{Oppenheimer-08} for
more details on this aspect.  We assume a \citet{Chabrier-03} initial
mass function throughout.

The primary distinguishing aspect of our code is the use of a highly
constrained heuristic model for galactic outflows.  The model we
follow is presented in \citet{Dave-13}, which utilises outflows
scalings expected for momentum-driven winds in sizeable galaxies
($\sigma>75\kms$), and energy-driven scalings in dwarf galaxies.
The galaxy velocity dispersion of $75\kms$ corresponds to a galaxy
baryonic (star + gas) mass of $M_{gal}\sim10^{10}\msolar$,
typically half of which is stars (although gas-poor satellites
can be almost entirely stars), at $z=0$. This mass is lower
at higher redshift given that it is a function of $H(z)^{-1}$, 
as described in \citet{Oppenheimer-08}.

In particular, we assume that the mass loading factor (i.e. the
mass outflow rate in units of the star formation rate) is
$\eta=150\kms/\sigma$ for galaxies with velocity dispersion
$\sigma>75\kms$, and $\eta= 75\times 150/\sigma^2$ for $\sigma<75\kms$.
Our previous results have generally favoured these latter scalings for
all galaxies, matching everything from intergalactic medium (IGM)
enrichment~\citep{Oppenheimer-Dave-06,Oppenheimer-12} to circumgalactic
gas properties~\citep{Ford-13} to galaxy mass-metallicity
relations~\citep{Finlator-Dave-08,Dave-Finlator-Oppenheimer-11} and
stellar mass functions~\citep{Dave-Oppenheimer-Finlator-11}.  However,
these prior simulations generally did not resolve dwarf galaxies
unlike our current simulation, and hence we found that
the steeper scaling of $\eta(\sigma)$ in this regime produced an
improved fit to the stellar mass function~\citep{Dave-13}.
Note that we always assume $v_{\mbox{w}}\propto\sigma$ for
all galaxies, consistent with
observations~\citep{Martin-Engelbracht-Gordon-05,Weiner-09}.
Interestingly, these scalings are similar to those produced
in the fully self-consistent outflow simulations of \citet{Hopkins-14},
the Feedback in Realistic Environments (FIRE) suite of zoom
simulations~\citep{Muratov-15}.  The highly simplified physical explanation is that
at low masses, supernova energy is sufficient to unbind a sizeable
fraction of the gas~\citep{Dekel-Silk-86}, but at high masses,
additional contributions from momentum input are necessary
\citep{Murray-Quataert-Thompson-10,Hopkins-Quataert-Murray-12}.

To implement outflows, we kinetically eject particles from the ISM
in order to mimic unfettered escape through ISM chimneys.  The mass
ejection rate is given by
\begin{equation}\label{outflow}
\dot{M}_{\mbox{\tiny wind}} = \eta \times \mbox{SFR},
\end{equation}
where the SFR is computed using a two-phase subgrid ISM model
\citep{Springel-Hernquist-03}.  If a gas particle has some probability
to form stars, it has $\eta$ times that probability to be ejected
in an outflow.  We then eject it in a direction given by {\bf v}$\times${\bf a},
with {\bf v} its instantaneous velocity and {\bf a} its instantaneous acceleration.
The velocity dispersion of the galaxy is estimated from an on-the-fly friends of friends galaxy finder;
see \citet{Oppenheimer-08} for details.  After this, we turn 
off hydrodynamic forces on the particle until it reaches a
density that is 10\% of the star-forming density threshold, i.e. 
$0.013\ cm^{-3}$, or if a timescale corresponding to 
$1.95 \times 10^{10}/(v_{\mbox{w}} \kms$) years has elapsed.

This simulation also includes a heuristic model to quench star
formation in massive galaxies tuned to reproduce the exponential
truncation of the stellar mass function.  Star formation in a given
galaxy is stopped depending on the quenching probability $P_Q$ given
in equation (\ref{quench}), which is a function of the velocity dispersion
$\sigma$ of the galaxy:
\begin{equation}\label{quench}
P_Q = 1 - \frac{1}{2} \erf{\frac{\log{\sigma}-\log{\sigma_{\rm med}}}{\log{\sigma_{\rm spread}}}}
\end{equation}
We use $\sigma_{\rm med}$=110$\kms$ as the velocity dispersion where
a galaxy has 50\% chance to have its star formation turned
off, and $\sigma_{\rm spread}$ = 32$\kms$ to describe some scatter
between $\sigma$ and the detailed physics of quenching.  Note that
this model does not attempt to directly model the physics of
quenching, it is only a way to reproduce the observed mass function
at the high mass end, and has virtually no effect below the knee
of the mass function. When a galaxy is chosen to stop its star
formation, any particle eligible for star formation first has its
quenching probability assessed, and if it is selected for quenching
then it is heated to 50 times the galaxy's virial temperature, which
unbinds it from the galaxy.  As discussed in \citet{Dave-13}, and as
we will demonstrate later, this quenching model does not substantially 
impact the overall \HI\ content of galaxies, though there are
some important environmental effects.  It is also important to note
that, owing to being tied to a threshold in $\sigma$, this quenching
prescription really only affects massive {\it central} galaxies;
the quenching of {\it satellite} galaxies occurs owing primarily
to tidal and hydrodynamic stripping effects that are self-consistently 
modeled in the simulation.

To examine dependencies on our feedback assumptions, we will use
another model without quenching, namely the ``vzw'' or momentum-driven
wind galactic outflow model.  This model, fully described in
\citet{Dave-Finlator-Oppenheimer-11}, has two main differences 
relative to the primary model used in this work.
First, the vzw model does not include the quenching
prescription as in our ezw model.  Second, the mass loading factor
$\eta$ is inversely proportional to the velocity dispersion $\sigma$
regardless of the size of the system.  Hence, we would only expect
a difference at the low $\sigma$ end ($\sigma<75\kms$) owing to the
ezw model adopting a steeper outflow scaling ($\eta\propto\sigma^{-2}$)
at these smaller $\sigma$.  For massive galaxies, we will use the comparison
between the two models to distinguish the effect of the star formation
quenching prescription, which only operates at high $\sigma$ where
the two models have the same mass loading factor $\eta$ dependency.

Our analysis will only consider galaxies that contain at least
64 star particles having a respective stellar mass of
1.45$\times$10$^{8}\msolar$. With this limit, the simulation produces
3,732 galaxies, among which 2,607 and 1,125 are central and satellite galaxies,
respectively. At the low masses, it was shown in \citet{Dave-13}
that galaxies are most \HI\ rich, with a gas content that is typically at
least as much as their stellar mass \citep[in accord
with observations: e.g.][]{Walter-08, Rupen-10}.
Given that a gas particle can spawn to 2 star particles\footnote{This
number is chosen as a compromise between desiring a gradual buildup
of stellar mass, versus not introducing too many new particles that
slows the computation.}, our stellar mass limit corresponds to 32
gas particle mass. This is probably suboptimal for fully
resolving ram pressure stripping processes, as the idealised tests of
ram pressure stripping in hot cluster gas by \citet{McCarthy-08} suggest
that significant non-convergence appears when the number of gas particles
drops below several hundred.  Hence our results for galaxies with
$M_*\la 10^9 M_\odot$ may be considered preliminary subject to more
detailed mass resolution convergence studies.  Nonetheless, the
results for our lowest mass galaxies tend to mostly lie on an
extrapolation from higher-mass galaxies, and hence the qualitative
trends we identify in this work are likely to be robust. Spatially,
even our smallest galaxies have a typical \HI-weighted radius of
$\sim 10$~kpc, well above our spatial resolution. The tests
for resolution convergence were mainly done by looking at the
differences between our high resolution simulation and another lower
resolution simulation with only $2\times 256^3$ particles, i.e. eight times 
less particles. We find that the two mass functions are very similar
above the mass resolution limit, which implies good convergence.
The \HI~richness versus stellar mass, however, has noticeable differences
between
the two simulations, with the high resolution simulation being
more \HI\ poor by $0.1-0.2$~dex. This discrepancy owes to the contributions of both an
increased stellar mass and a decreased \HI~mass for a given halo
mass at higher resolution, since higher resolution simulations are
more efficient in cooling the gas and hence in forming stars.
Despite these relatively minor differences, the resolution convergence
is sufficient not to affect our main conclusions. For more detailed
tests of resolution convergence, we refer the reader to \citet{Dave-13}.

Each halo is identified using a spherical overdensity (SO) algorithm
\citep{Keres-05}, which takes each galaxy centre and expands a sphere
around it to enclose an average overdensity given by
% \citep{Oppenheimer-Dave-08}
\begin{equation}
\delta = 6\pi^2 (1 + 0.4093(1/f_\Omega  -1)^{0.9052}) - 1
\end{equation}
with 
\begin{equation}
f_\Omega = {\Omega_m(1+z)^3\over {\Omega_m(1+z)^3 + (1-\Omega_m-\Omega_\Lambda)(1+z)^2 + \Omega_\Lambda}}.
\end{equation}
This corresponds to an overdensity of $\delta \simeq 110$ at $z=0$.
In our simulation volume, the most massive halo
at redshift $z=0$ is $M_{halo}$ = $1.32\times10^{14}\msolar$; hence
our environmental study focuses on the field to moderate-sized group
environment.

%---------------------------------------------------------------------------------------------------------------------------
\subsection{Computing the {\HI} content}\label{simu2}
\indent

We identify galaxies using Spline Kernel Interpolative
Denmax\footnote{http://www-hpcc.astro.washington.edu/tools/skid.html}
({\sc skid}) as bound collections of stars and star-forming gas.  We associate
each galaxy with a halo identified using a spherical overdensity
algorithm, as described above.  The central
galaxy is taken to be the most massive galaxy in the halo, and we consider
all the others as satellites.  While SKID adequately captures the
stellar and molecular mass, significant amounts of \HI\ can be
present outside of the star-forming gas. We compute the \HI\ content 
of the simulated galaxies following the methodology described 
in full detail in \citet{Dave-13}; we review the main points here
as well as modifications to the previous modeling.

First, we consider all particles within a sphere of a radius given
by the outermost particle in the {\sc skid} galaxy, and associate all the
\HI\ to that galaxy.  For close pairs of galaxies, we assign the particles
to the galaxy with higher gravitational force to avoid double-counting
\HI.  For each gas particle, we calculate the neutral fraction based on the
assumption that the given particle is a sphere with a kernel
density profile given by the SPH kernel, and that it is bathed in 
radiation from the metagalactic radiation field given by
\citet{Haardt-Madau-01}.  We then integrate the column density inwards
from the particle surface until it reaches a threshold column density
where the particle becomes sufficiently neutral.  The mass fraction
within this radius is then considered to be 90\% neutral (since some
mass in the outskirts of galaxies remains ionised), while
the mass outside this radius has the optically-thin ionisation 
fraction.  We refer to this as the ``auto-shielding" approximation.
and \citet{Dave-13} showed that it yields good agreement
with the full radiative transfer simulations of \citet{Faucher-Giguere-10}.
We tested the same method by using a sphere of radius 
1.5$\times$ higher than that given by the outermost particle in the SKID
galaxy and ended up with a $\la 5\%$ typical increase in the total
\HI~mass.

Compared to \citet{Dave-13}, we have changed the column density
threshold within which we consider the gas to be fully neutral in
the auto-shielding calculation.  \citet{Dave-13} utilised
$\NHI=10^{17.2}\cms$, which corresponds to an optical depth of unity
to ionising radiation.  However, given that the ionisation level
of an optically-thin gas bathed in metagalactic flux is much smaller
than unity, typically $10^{-5}-10^{-6}$~\citep[e.g.][]{Dave-10},
an optical depth of unity will still not produce a substantial \HI\
fraction.  Instead, we need an optical depth that will result in
an \HI\ fraction of 0.5.  If we assume that the ambient gas has an
ionisation fraction of $10^{-5}$, we require that the optical depth
be 10.8 instead of unity.  The resulting column density threshold
thus rises from $1.6\times 10^{17}\cms$ to $1.7\times 10^{18}\cms$.
We, therefore, use this latter column density threshold to
calculate the radius within which the gas is assumed to be fully
neutral. We have checked that the agreement with the radiative transfer simulations
of \citet{Faucher-Giguere-10} remains acceptable, and in fact is slightly improved.  In \S\ref{himf} we will compare our results to that using
the old threshold value.  If such a radius does not exist (typically
for particles with density $n\la 10^{-2}\cmc$), then the particle
is fully optically thin and its ionisation fraction is calculated
assuming ionisation balance with the metagalactic flux.

Next, we must also determine how much of this shielded gas is
molecular.  For this, we employ the $H_2$ formation model of
\citet{Krumholz-Gnedin-11}, which we compute for each particle.
They consider radiative transfer coupled with the formation and
dissociation balance of the molecular hydrogen in a steady state
molecular cloud. The resulting $H_2$ fraction is given by
\begin{equation}\label{kg11} 
f_{H_2}\simeq 1 -
\left(\frac{3}{4}\right)\frac{s}{1+0.25\times s}
\end{equation} 
where $s$ is a term dependent on the formation rate of the molecular hydrogen
from dust grains (which depends on metallicity), the dust cross
section per H nucleus, and the ambient intensity of the ultra-violet
radiation field.  Note that in \citet{Dave-13} the default model
was the observationally-constrained pressure law of \citet{Leroy-08}.
We use \citet{Krumholz-Gnedin-11} here because we want to study the
evolution of \HI\ to higher redshifts, where the metallicities can
significantly deviate from the solar abundances typical in the
\citet{Leroy-08} data; the \citet{Krumholz-Gnedin-11} model accounts
for such metallicity dependencies in $H_2$ formation. 
\citet{Dave-13} demonstrated that at $z=0$ this makes essentially no 
difference to most \HI\ properties, but at higher redshifts there are
more substantial differences. Briefly, high redshift galaxies
have lower metallicities, resulting in less molecular gas and hence an increased
\HI~mass fraction.  \citet{Dave-13} showed that at $z=3$ the HIMF using the
\citet{Krumholz-Gnedin-11} model results in a mass function that is about two
times higher than that derived using the \citet{Leroy-08} prescription.
Nonetheless, in this work we mostly focus on lower redshifts where the
differences are minimal.

With this prescription, we separate each gas particle into a neutral,
molecular, and ionised component.  The \HI\ mass of the galaxy is
then the sum of the neutral mass in all its associated particles,
and likewise for the molecular mass.

%---------------------------------------------------------------------------------------------------------------------------
\subsection{Tracking progenitors}\label{simu3}
\indent

In this paper, we will be interested in tracking the evolution of
\HI\ in individual galaxies back in cosmic time.  We do this by
associating a given galaxy at $z=0$ to its most massive progenitor
at all of our previous 134 outputs back to $z=30$.  We define the
most massive progenitor as the galaxy (identified by {\sc
skid}) at an earlier output containing the largest number of star
particles in common with the galaxy at $z=0$.  We choose the
limiting number of particles for a group to be considered
as the galaxy progenitor to be 32 particles, which
is below the resolution limit.  This choice does not affect the
resolved progenitor history, but allows the history to be extended
somewhat further back in time.

From this, we can identify major mergers by a simple prescription,
following \citet{Gabor-Dave-12}.  For an assumed major merger ratio
$r$, we search the stellar mass growth history for jumps in excess
of $1/(1+r)$. The minimum value of $r$ for which continual infall 
and a merger could be distinguished depends on the redshift 
and the time between successive outputs, but it is 
generally quite far below the minimum ratio of $1:3$  that we
use in the analysis, i.e. our outputs 
are frequent enough, every 100 to 300 Myr, such that no 
galaxy would ever grow by $33\%$ just from \textit{in situ} star formation 
between outputs alone. There are, however, occasional complications
with this approach.
First, a star particle might, at a given output, be attributed to no group.
Second, two groups brought close along their path (without merging)
could be regarded as one group by {\sc skid}, but would later
``unmerge".  To address the first problem, we look back in time until
the star belongs to a galaxy, and then attribute that star to the
descendent of that galaxy.  To address the second problem, we check
which galaxies at the previous output are composed of two separate
galaxies. We track those merging units until every star in both
galaxies is assigned to a single galaxy.  From then on, we assign
the stars located in the smaller group to that single galaxy.  We
have found that this robustly identifies merger events despite the
dynamical nature of the merger encounter.

%---------------------------------------------------------------------------------------------------------------------------
\section{\HI\ and environment at \lowercase {z=0}}\label{hi}

%---------------------------------------------------------------------------------------------------------------------------
\subsection{\HI\ mass function}\label{himf}
\indent

The most basic statistical property of \HI\ galaxies is the \HI\
mass function (HIMF).  \citet{Dave-13} showed that the simulation
we use here, together with their prescription to compute the \HI\
content, yields an \HI\ mass function that is in good agreement
with observations down to the lowest resolvable masses ($\sim 10^8
M_\odot$).  This is a non-trivial success that has been difficult
to achieve in simulations. Recent SAMs have done better but still
show an excess at $\MHI\sim 10^9 M_\odot$ \citep{Lagos-14}.
In this section we test our HIMF
in more detail, in particular separating it into bins of stellar
mass to compare with recent data from \citet{Lemonias-13}.

%________________________________________________
\begin{figure*}
 \begin{center}
 \includegraphics[scale=0.6]{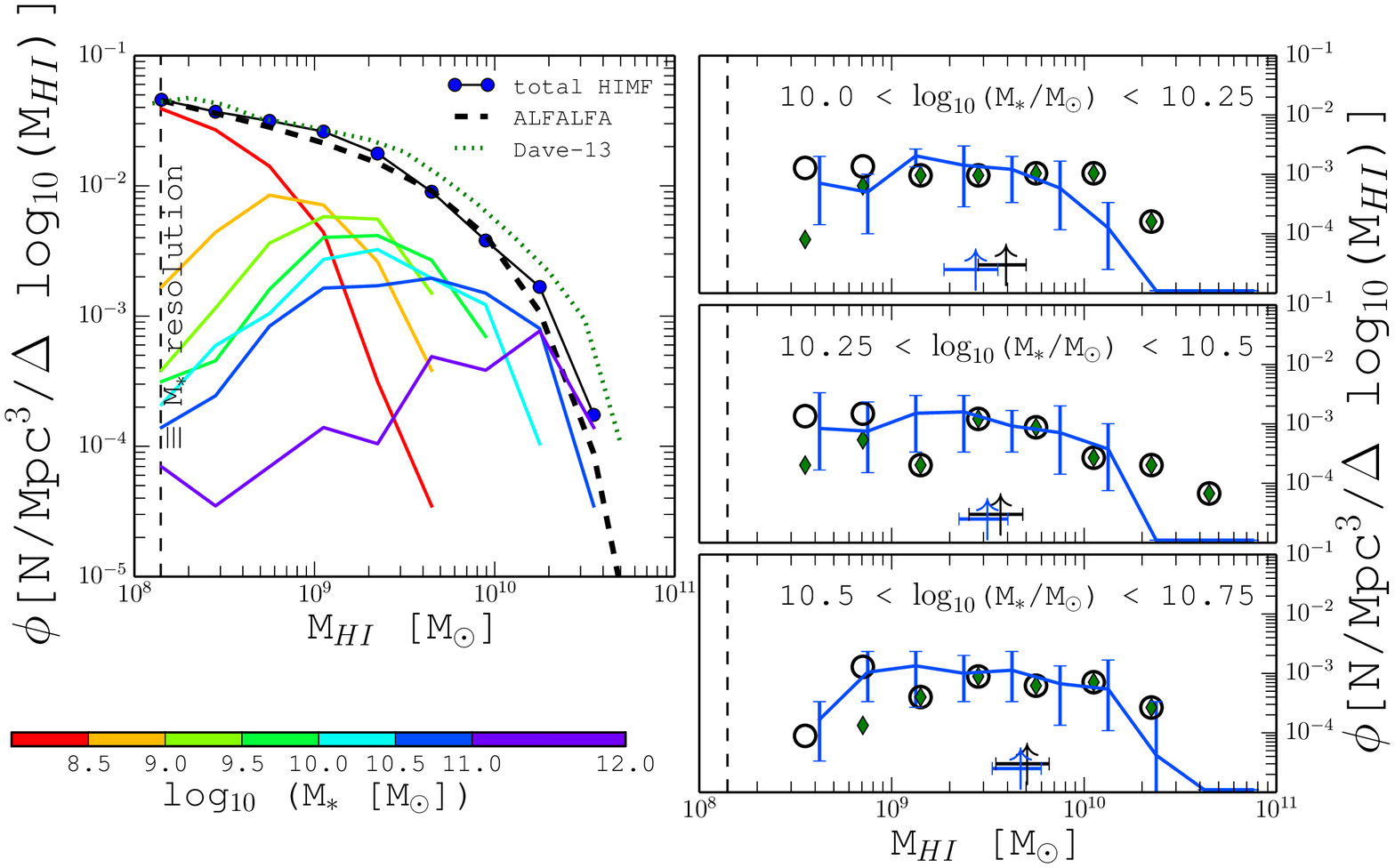}
 \caption{The \HI\ mass function at $z=0$ for the population
 of galaxies generated by the simulation described in section \ref{simu}.
 \textit{Left panel}: We plot our total HIMF as the black line (with blue circles),
 the simulated HIMF of \citet{Dave-13} as the dotted green line,
 and the HIMF of galaxies from the ALFALFA \HI\ survey~\citep{Haynes-11} as the black dashed line.
 Contributions to our simulated HIMF from bins of 0.5 dex in stellar mass
 are indicated by the colored lines shown by the color-bar
 for the mass range $M_*$ = $10^8$ - $10^{12}\msolar$
 (the highest mass bin covers 1 dex owing to the small number
 of galaxies in this mass range).
 The vertical dashed line shows the stellar mass resolution limit for reference; 
 there is no formal \HI\ mass resolution limit, but Figure~\ref{mhimstar} shows
 that the \HI\ and stellar masses are comparable in galaxies of this mass.
 \textit{Right panel}: Comparison between our simulated galaxy HIMF
 (blue lines; where error bars indicate the cosmic variance in each $\MHI\ $ bin)
 and the observational data of \citet{Lemonias-13} (green diamonds: \HI~detected sample;
 open black circles: full sample with non-HI-detected galaxies included
 at their upper limits) for three different bins in stellar mass. The upward arrows on 
 the bottom of each right panel indicate the weighted average of 
 the HI mass with errors computed as described in the text (black: observed sample;
 blue: simulation prediction).}
\label{fighimf}
\end{center}
\end{figure*}

Figure \ref{fighimf} shows the HIMF of our galaxies (black
line with blue circles), down to an \HI~ mass similar
to the stellar mass resolution limit, namely $\MHI=1.45\times 10^8
M_\odot$.  Note that our simulation does not have a formal \HI~mass
resolution limit, only a stellar mass resolution limit.  Here we
leverage the fact that, near this resolution limit, the \HI\ and
stellar masses are typically comparable, and hence we quote our
HIMF down to this mass limit in \HI\ as well.  However, this means
that we may miss some galaxies with large \HI\ fractions below our
stellar resolution limit.  We found that reducing the stellar mass
limit by a factor of two results in only a $\sim15\%$ increase in
the \HI\ mass function at $M_{\rm HI}\approx 1.4\times 10^8 M_\odot$.
This suggests that our predicted HIMF faint end is not very sensitive to
our stellar mass cut excluding low-$M_*$, large $M_{\rm HI}$
galaxies.

Since our $\NHI$ self-shielding threshold is altered from that
assumed in \citet{Dave-13}, we compare our HIMF to theirs shown as
the dotted green line.  Overall, this change is consistent with a
$\sim 30$\% shift towards lower $\MHI$ (i.e.  leftwards) versus the
\citet{Dave-13} HIMF, since it results in a lower fraction of each
gas particle being neutral.  Consequently, this produces a larger
change in the amplitude at the high-mass end.  Our new HIMF is in better
agreement with the observed HIMF from the ALFALFA survey~\citep[dashed
line;][]{Haynes-11}, modulo the concerns discussed above
regarding missing low-$M_*$ galaxies with very high $M_{\rm HI}$.

\begin{figure}
 \begin{center}
 \includegraphics[scale=0.5]{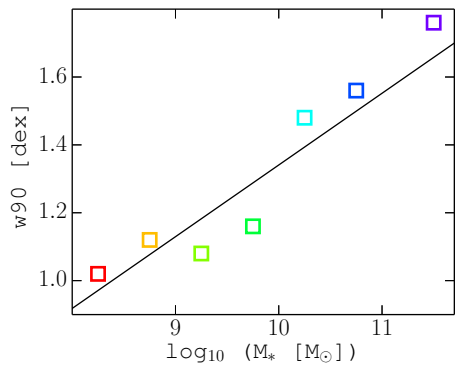}
 \caption{The mass range $\Delta \log \MHI$ 
enclosing 90\% of the galaxies in the HIMF ({\sc w90}) versus the mid-value
of the $M_*$ bin. Each stellar mass-binned HIMF shown
in Figure \ref{fighimf} is fit with a Gaussian and then
{\sc w90} is computed. The colour scheme is the same as that 
in Figure \ref{fighimf}. {\sc w90} steadily increases from
1.03 dex at the lowest $M_*$ bin to 1.76 dex at the highest
$M_*$ bin. The black line shows a log-linear fit with a slope
$0.23$.}
\label{figw90}
\end{center}
\end{figure}

We further divide our HIMF into bins of 0.5~dex in stellar mass $M_*$,
shown as the coloured lines, from $M_*=10^8\msolar-10^{12}\msolar$.  
This shows how the HIMF is comprised of galaxies with different stellar masses. 
Since $\MHI$ correlates with $M_*$, there is a general trend that the
HIMF shifts towards higher $\MHI$ at higher $M_*$.  There is also a
slight trend for an increased width of the HIMF as one moves towards
higher $M_*$, reflective of the fact that low-$M_*$ galaxies have
more uniformly high \HI\ richness, while more massive galaxies can
have a wider range of \HI\ content. To quantify this wider range in $\MHI$ 
at higher $M_*$, we perform Gaussian fits to the HIMF 
for each stellar mass bin.  We compute the mass range 
$\Delta\log\MHI$ centred at the mean value of the Gaussian fit 
that encloses 90\% of the galaxies in each $M_*$ bin, 
which is indicated as the ``{\sc w90}" value. We then plot
``{\sc w90}" versus $M_*$ in Figure \ref{figw90}, with the same
colour scheme as in Figure \ref{fighimf}. 
{\sc w90} steadily increases from 1.03~dex for the lowest $M_*$ bin to
1.76~dex for the highest $M_*$ bin, roughly following the relation
$\Delta\log(\MHI/M_\odot) \simeq0.23\log (M_*/M_\odot)-0.9$.

If there was perfect correlation between $\MHI$ and $M_*$, then the
HIMFs separated into $M_*$ bins would have no overlap.  Hence, the fact
that the HIMF spreads over a larger range than the bin size in
$\MHI$ is a measure of the scatter between $M_*$ and $\MHI$.  This
scatter is a key indicator of how the \HI\ content of galaxies
varies with environmental influences.  For instance, \citet{Moran-12}
showed that galaxies with enhanced total $\MHI$ tend to have the
excess \HI\ in their outskirts, accompanied by a metallicity drop,
indicative of recent accretion, although \citet{Wang-14} find
that roughly half of \HI\ rich galaxies have the excess predominantly
in their centre, and not in the outskirts.  One possible physical
interpretation is that the
spread between $\MHI$ and $M_*$ tracks the frequency and amount of
recent accretion (or lack thereof) that gives rise to variations
in \HI\ without accompanying immediate variations in $M_*$.

To test this aspect of our simulations, we can compare our results
to the data from the GASS survey\footnote{\label{gassnote}http://www.mpa-garching.mpg.de/GASS/}
separated into stellar mass bins by \citet{Lemonias-13}.
The GASS survey is stellar mass-selected
down to $M_*= 10^{10}M_\odot$, and hence provides a fair
comparison to our stellar mass-selected simulated galaxy sample,
at least down to their completeness limit.

The right panels of Figure~\ref{fighimf} show the comparison to the
data from \citet{Lemonias-13}, shown as the green diamonds for the
\HI-detected galaxies only, and as black circles for all galaxies
including those without detected \HI\ placed at their upper limit in \HI,
in three bins with widths of 0.25~dex in $M_*$.  For a fairer comparison,
we similarly adjusted the \HI~mass of the simulated galaxies according
to the GASS data limits used in \citet{Lemonias-13}, namely we set
a minimum mass of $M_{HI}/M_* = 1.5\%$ for galaxies with $\log_{10}
(M_*/M_\odot)>10.5$ and $\log_{10} (M_{HI}/M_\odot) = 8.7$ for
galaxies with $\log_{10} (M_*/M_\odot) < 10.5$.  Hence the simulation
points should most directly be compared to the open black circles;
the green diamonds are shown so the reader can assess the
contribution from non-detected galaxies to the observed sample.
Error bars on the HIMF indicate the cosmic variance in each $M_{HI}$
bin; they are computed as the scatter of the HIMFs amongst the
octants of the cubical simulation volume.

At low \HI\ masses, the observation data (black circles) fall within
the simulation predictions' (blue line) cosmic variance.  At high
masses, however, our simulated galaxy population does not extend
up to the largest \HI\ galaxies, showing that our model does not
produce galaxies at M$_*\ga 10^{10}\msolar$ with very high
\HI\ content. This may partly owe to our quenching prescription
that, in some cases, starts to affects galaxies at these stellar
masses;  we will demonstrate later that once galaxies enter the
mass regime where our quenching starts to become important,
our simulation does not reproduce the observations quite as well,
suggesting that our quenching model is not fully compatible with
the \HI\ data.  On the other hand, the agreement is if
anything somewhat better at the highest $M_*$ values, suggesting
perhaps that this is a more fundamental issue with our simulation.

Near the bottom of the right panels of Figure~\ref{fighimf}, we
show the weighted average of the \HI~mass in each stellar mass bin
as the upward arrows.  We use the normalised number density of
galaxies ($\phi_i/\sum \phi_i $) to determine the central value of each
\HI~mass bin. The errors are also weighted by the normalised number
density of galaxies, taking the error for each bin as the bin width
in \HI~mass.  The colours correspond to the various data points;
we only show the values for the data and models with \HI-undetected
galaxies set at the detection limit (black and blue upward arrows,
respectively).  The data are more consistent with our simulations
in the higher stellar mass bins. The differences at low masses
mainly owe to the fewer high \HI~mass galaxies in the lower stellar
mass bin, and for this reason, the weighted \HI\ mean is lower for
the simulated galaxies. The difference between the means decreases
from $\sim40\%$ to $\sim10\%$ in the highest stellar mass bin.

\begin{figure}
 \begin{center}
 \includegraphics[scale=0.6]{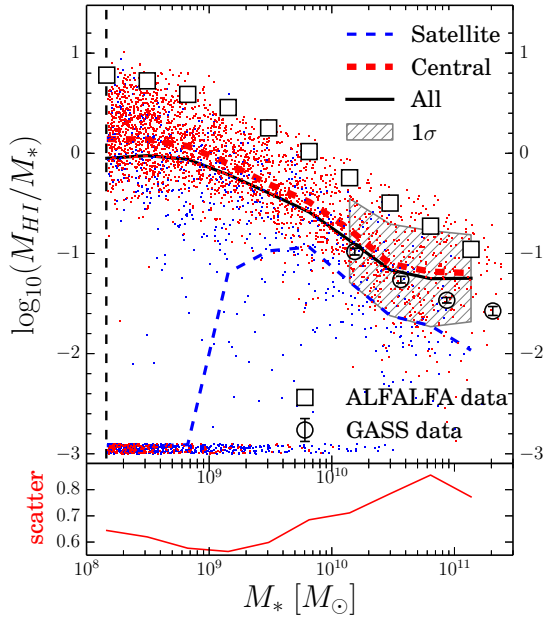}
 \caption{The \HI~ richness ($M_{HI}/M_*$) vs. $M_*$ of $z=0$ galaxies from 
simulations and observations. Red points are central galaxies, blue points
are satellites. The galaxies grouped near the bottom of the plots contain
galaxies with $M_{HI}/M_*\in$ [$0$, $10^{-3}$], with their distribution peaking at
$\sim10^{-6}$. The black line is the running median of
the simulated galaxy \HI~richness, with the $1\sigma$ uncertainty indicated by
the grey hatched area on the massive end (overlapping with GASS).
The data from GASS DR3 \citep{Catinella-13}
are shown as the empty circles with
error bars representing $1\sigma$ uncertainties of the medians. The non-detected
sources in GASS are included at their upper limit.
ALFALFA data \citep{Huang-12} are presented as black squares.
The striking difference between ALFALFA and GASS owes
to the selection criteria, as described in \citet{Catinella-10}.
The stellar mass selection of GASS is more directly comparable to our
simulated sample, which is consistent with GASS observations within the
uncertainties.  The lower panel is the log amplitude
of the scatter around the central galaxies; the scatter generally
increases with mass. 
}

 \label{mhimstar}
\end{center}
\end{figure}

This comparison suggests that our simulations are generally reproducing
the scatter in the relationship between $\MHI$ and $M_*$, at least
down to $M_*\approx 10^{10}M_\odot$.  To quantify this further, we
plot in Figure~\ref{mhimstar} \HI\ richness ($\MHI/M_*$)
as a function of $M_*$. We subdivide it into central (red) and
satellite (blue) galaxies.
The \HI-poor galaxies dominate the satellite population
especially at low stellar masses, while the central galaxies
drive the shape of the relationship between \HI-richness
and $M_*$ because they are more numerous.
In other words, a similarity can be seen between
the black line and the red dashed line, where the maximum
difference is $\sim60\%$ at M$_* = 1.45\times10^{8}\msolar$.
We will discuss the satellite population further in \S\ref{csHI}.
The grey hatched area represents the $1\sigma$
uncertainty of the simulated sample falling in the same range of stellar
mass as the GASS data. This is identical to Figure~4 in \citet{Dave-13},
except that we are now employing our new $N_{\rm HI}$
self-shielding limit as described in \S\ref{simu2}.

We show the median of the GASS data (open circles) from
their third data release~\citep{Catinella-13} with the error bars
representing the $1\sigma$ uncertainty in the median computed using
jackknife resampling over 8 subsamples.  The simulations
and the observations are consistent at the $\la 2\sigma$
level at lower masses, but at higher stellar masses ($M_*\ga 10^{10.5}$) we
see that the simulation (black line) deviates from the
observations (open circles).  This deviation owes to our quenching model,
which stops the stellar mass growth, but the galaxies can still
accrete some gas. Qualitatively, this suggests that our 
quenching model needs to be more effective at preventing cold gas accretion.
We also show the results from the ALFALFA
survey \citep[black squares;][]{Huang-12} for reference,
but these are not for direct comparison to our models.  These values
are higher than those seen in the GASS survey since ALFALFA is
\HI-selected and hence picks out higher-$\MHI$ galaxies at any given
mass.  To compare to ALFALFA we would have to mimic the volume and
selection function of that survey, which we leave for future
work\footnote{We note that this has caused some confusion in the
literature.  For instance, the Illustris simulation \citep{Vogelsberger-14}
agrees well with the ALFALFA \HI\ richness data, but this is using
all their galaxies without mimicking the ALFALFA selection.}.  Hence,
while the ALFALFA HIMF is appropriate to compare to our models
(under the reasonable assumption that ALFALFA has appropriately
accounted for its selection volume), the \HI\ richness from that
survey is biased high~\citep{Catinella-10}.  Thus our predicted
\HI\ richness should not be directly compared to the ALFALFA data.
Nonetheless, it is interesting that our models follow the
shape of the ALFALFA \HI~richness vs M$_*$ to low masses, offset by a fixed
factor that is similar to the offset between GASS and ALFALFA in
their overlapping $M_*$ range.

In summary, our simulated HIMF agrees very well with ALFALFA
observations, particularly with our more physically-motivated
criterion for the column density where auto-self-shielding becomes important.
Binning the HIMF into $M_*$ bins, for non-quenched galaxies, our
simulation agrees well with observations from the GASS survey,
suggesting that the scatter in $\MHI$ versus  $M_*$ also agrees
with observations, although the simulations do not produce the most
\HI-rich galaxies at lower stellar masses.  For quenched galaxies (or galaxies
on the verge of quenching), we likely require a more sophisticated
quenching prescription to match the data at the low-$\MHI$ end, and
this could potentially provide a new and interesting constraint to test our
ongoing improvements to our quenching model~\citep[e.g.][]{Gabor-Dave-15}.
This is also evident in 
our \HI\ richness predictions, which match well with stellar mass-selected
observations up to the masses where once again our quenching model might
be inadequate. Modulo these small discrepancies, the general agreement
in the scatter between $\MHI$ and $M_*$ is encouraging and the first time it
has been demonstrated for cosmologically-based \HI\ models of any type.

%---------------------------------------------------------------------------------------------------------------------------
\subsection{Satellite galaxies}\label{csHI}

Once a galaxy falls into another galaxy's halo, a number of physical
processes can act to remove or deplete its gas.  These include tidal
stripping, ram pressure
stripping~\citep[e.g.][]{Gunn-Gott-72,vanGorkom-03}, viscous
stripping~\citep[e.g.][]{Marcolini-Brighenti-D'Ercole-03}, induced
star formation from encounters with other
galaxies~\citep[harassment;][]{Moore-96}, and starvation
of inflow to sustain star formation.  Hence the behaviour
of satellites within a halo is expected to differ from that of the
central galaxy.  These various physical processes are expected to
leave different signatures on the \HI\ content, based on how much
ambient gas is within the central galaxy's halo.
%{\red For instance,
%tidal stripping is independent of the halo gas, being driven primarily
%by encounters with other galaxies \citep{Willman-04}, whereas 
%satellite galaxies are affected by tidal forces due to interaction with a
%nearby engulfing halo.}
Ram pressure stripping scales as 
$\propto \rho v^2$~\citep{Gunn-Gott-72},
where $\rho$ is the ambient gas density and
$v$ is the relative velocity of the satellite, and hence is effective
in high velocity dispersion halos with
significant halo gas.  Induced star formation
relies on the presence of other satellites to stimulate a starburst
\citep[from tidal interactions;][]{Li-08}.
Starvation in satellites occurs when its own hot gaseous halo gets
stripped, and its feeding streams have been disconnected since
they now head towards the central galaxy.
The \HI\ gas, generally being more loosely bound and hence more
easily susceptible to stripping and heating than the molecular gas
and stars, provides a unique and interesting test bed to study the
importance of these environmental processes over a wide range of
halo masses.  In this section we focus on the properties of centrals
and satellites as a function of halo mass.

\begin{figure}
 \begin{center}
 \includegraphics[scale=0.8]{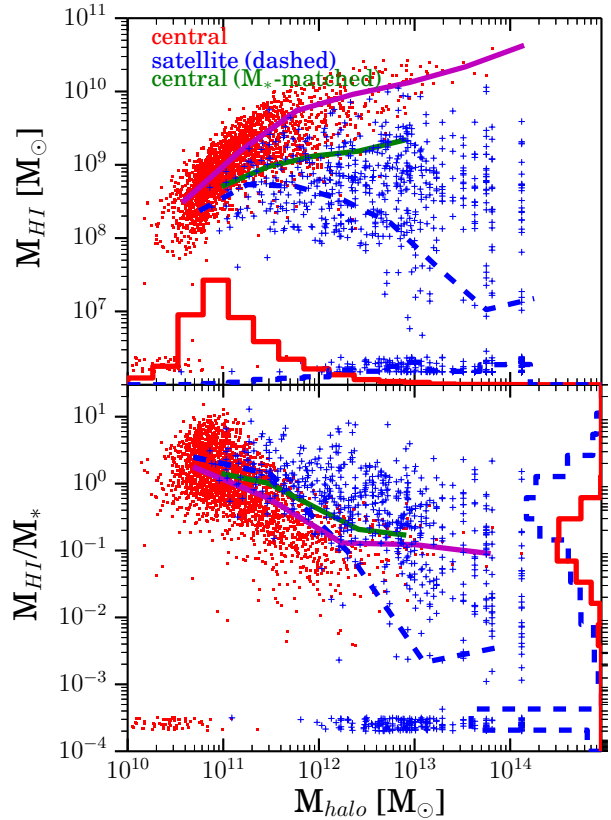}
 \caption{The \HI\ mass (top) and \HI\ richness (bottom)
 of simulated galaxies as a function of halo mass at $z=0$.
 Centrals are shown in red (points), satellites in blue (crosses),
 and the magenta lines and blue dashed lines correspond to 
 their running median values respectively. The green lines
 show the median values of the central galaxies having
 stellar masses within the stellar mass range of the satellite galaxies
 in the same halo mass bin (``$M_*$-matched" centrals).
 Histograms in the upper panel show the distribution
 of halo masses for all central (red line) and satellite (blue dashed line) galaxies.
 Histograms in the lower panel show the distribution
 of \HI\ richness for centrals (red line) and satellites (blue dashed line)
 with $M_{halo}>$  10$^{12}\msolar$.
 Environmental effects become very prominent for satellite galaxies
 with $M_{halo}\geq10^{12}\msolar$.}
 \label{figcensat}
\end{center}
\end{figure}

Figure \ref{figcensat} shows the \HI\ mass (upper panel) and \HI\
richness (lower panel) as a function of halo mass for galaxies in
our simulations.  Central galaxies are shown as red points, and satellites
are indicated by blue crosses.  The data points along the bottom
of each panel correspond to galaxies with \HI~mass
(or \HI~richness) below the y-axis limit, and they are given
some artificial scatter for visibility.
The magenta lines and blue dashed lines show 
the running median values for centrals and satellites,
respectively. The green lines show the median values
for the central galaxies within the stellar mass
range of the satellite galaxies in the same halo mass bin;
this specifically isolates the difference between centrals and
satellites owing to halo mass, in a stellar mass-matched sample.
Histograms in the upper panel (red: centrals; blue
dashed: satellites) show the distribution of all resolved
galaxies (M$_*\geq1.45\times10^8\msolar$) in terms of halo mass.
Histograms in the bottom panel indicate the
distribution of galaxies with $M_{halo}>10^{12}\msolar$
(along with our resolution limit of $\resgal$).  The amplitudes of the
histograms in each panel are arbitrary, as these are purely for comparing
the abundance of centrals versus satellites.

The top panel shows that the \HI\ mass of centrals 
increases steadily with halo mass, though there is a marked transition 
at $M_{\rm halo}\ga 10^{12}M_\odot$ above which the relationship is less steep.  
This trend mimics the drop in sSFR around this halo
mass~\citep[e.g.][]{Dave-Oppenheimer-Finlator-11,Salim-07}, which arises owing to a
combination of the increased presence of a hot gaseous halo along
with quenching feedback.  Meanwhile, the \HI\ mass in satellites
is roughly independent of halo mass, rising with centrals at low masses
but decreasing in more massive halos.  The histogram along the bottom
shows that satellites strongly increase in relative abundance at
larger halo masses, as expected from halo occupation distribution
statistics~\citep[e.g.][]{Berlind-03}.
In addition, the upper panel also shows the median value of the \HI~
masses of the central galaxies having the same range of stellar mass
as the satellites in the same halo mass bin (green line). We can
only compute this quantity for $M_{halo}\la10^{13}\msolar$,
because we chose to select central galaxies within $\sim2\sigma$ of the
satellite stellar mass range, and the central 
galaxy stellar masses lay outside of this range in the highest halo mass bins.
Despite this limitation, we can still see that the effect of halo mass 
on the gas content of satellites is still visible and seems to become more
prominent beyond  $M_{halo}\simeq10^{12}\msolar$.

\begin{figure}
 \begin{center}
 \includegraphics[scale=0.45]{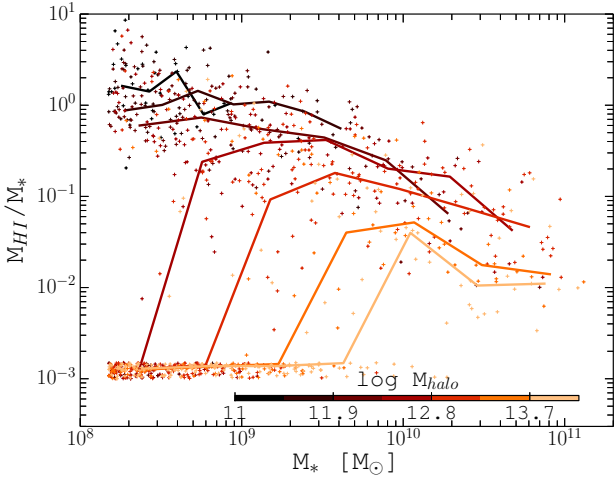}
 \caption{The \HI\ richness of satellite galaxies versus
 their stellar mass at $z=0$. The galaxies are colour-coded 
 by their respective halo mass as shown in the colour bar.
 The lines represent the median values for each halo mass bin.
 The points at the bottom of the figure show galaxies with \HI\
 richness below $10^{-3}$.
 This plot shows that the host halo mass 
 starts to influence the majority of satellites at $M_{\rm halo} \ga 10^{12}\msolar$.}
 \label{figsathalo}
\end{center}
\end{figure}

For {\HI} richness (Figure \ref{figcensat}, bottom panel),
the centrals and the satellites follow a similar trend in the low mass regime,
with the satellites having slightly higher \HI\ richness owing to their
typically lower $M_*$.  Hence at $M_{\rm halo}\la 10^{12}\msolar$,
it seems that satellites' \HI\ content is essentially unaffected
by being within the halo of another more massive galaxy.

The situation is different for $M_{\rm halo}\ga
10^{12}\msolar$.  Here, the median satellites \HI\ richness drops
dramatically, and even though their masses are lower (and hence by
the overall correlation should have higher \HI\ richness), their
\HI\ richness is significantly below that of centrals.  Looking
more closely, this drop is driven by a rapidly growing population
of satellites with essentially no \HI, whereas the satellites that
have \HI\ tend to lie slightly above the centrals following the
trend at lower halo masses.  The blue histogram (dashed) along the
right axis shows this bimodality for galaxies with $M_{\rm halo}>
10^{12}\msolar$.  This strong bimodality, analogous to that in a
galaxy colour-magnitude diagram, indicates that the \HI\ in satellites
is fairly rapidly removed once it enters a massive halo.
While the lower peak is enhanced in amplitude since we have
placed all \HI-poor galaxies at $\sim2\times10^{-4}$, nonetheless there is clearly an
emerging population of low-\HI\ galaxies that is distinct from the
extrapolated trend from lower mass halos, regardless of the binning.

Note that we only considered $\resgal$ galaxies, which could in
principle underestimate the median \HI~richness of satellites if
this mass resolution limit excludes a population of lower stellar
mass satellites with very high $M_{HI}/M_*$. However, for this to
happen, there would suddenly have to be many \HI-rich galaxies in
massive halos just below our $M_*$ resolution limit.  This seems
rather unlikely, given that in Figure~\ref{figsathalo} we will show
that virtually all low-mass satellites in massive halos are \HI-poor.
Also note that there are some \HI-poor central galaxies with
low halo masses.  As discussed and quantified in \citet{Gabor-Dave-15},
these are likely satellite galaxies whose orbits take them just
outside their massive host halo, and thus are identified as centrals
by our spherical overdensity halo finder.  The contribution of these
galaxies is minor, but had we identified these as satellites of
massive halos, this would strengthen the bimodality.  We will examine
the role of satellite orbits in stripping and bimodality in a future
paper.

We show in Figure \ref{figsathalo} the relation between
\HI\ richness and stellar mass for satellite galaxies within different
halo mass bins differentiated by their colours: from black for
$\log$(M$_{halo}\msolar)\sim$ 11 to yellow for $\log$(M$_{halo}\msolar)\sim$ 14.  From
the figure, we can see that  starting from $M_{\rm halo}\sim10^{12}\msolar$
(as mentioned before), there is a clear evidence for a growing population of gas-poor
satellite galaxies. The \HI-poor satellite population (shown by the
data points near the bottom of the plot) is dominated by galaxies located in more
massive halos.  Small satellites (low stellar mass) get their gas
content easily stripped owing to less gravitational pull binding the
gas to the galaxy, leading to the higher number of galaxies
at lower stellar masses in the highest halo mass bins. This is
an interesting prediction, but observation are not yet capable of
detecting the \HI\ content of such small galaxies in dense environments
expected to be dominated by hot halo gas. For instance, the upper limits in
\citet{Cortese-11} are well above what we predict for \HI~content.
Meanwhile, the {\sc Little
Things}\footnote{https://science.nrao.edu/science/surveys/littlethings}
survey observes only very nearby dwarf galaxies, which do not lie in 
sufficiently massive halos.

The oft-noted emergence of a hot gaseous halo around that mass
scale~\citep[e.g.][]{Keres-05,Gabor-Dave-12} hints at a connection
between the presence of a hot halo and the removal of \HI, which
favours a removal mechanism associated with gas stripping.  However,
there are also many more satellite galaxies in such massive halos
that could tidally strip or harass satellites, so the connection
is not completely certain.  Nonetheless, as shown in \citet{Cunnama-14}
from hydrodynamic simulations, \HI-poor satellites generally
feel a ram pressure force sufficient to remove its gas whereas \HI-rich
ones do not.  This suggests a strong link between ram pressure stripping
and satellite \HI\ removal, which our results corroborate from a 
different perspective. We note that halos start to have very small 
amount of hot gas starting with halo masses of $\sim10^{11}\msolar$ from which some environmental 
effect might slightly be seen (which is the case on Fig. \ref{figsathalo})

\begin{figure}
 \begin{center}
 \includegraphics[scale=0.6]{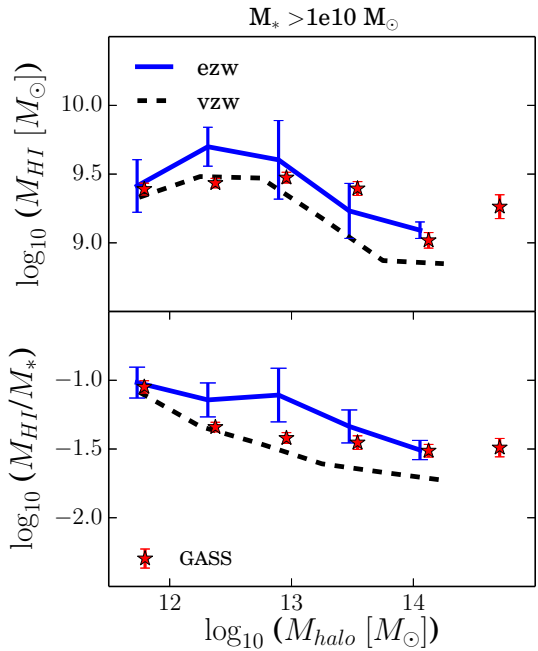}
 \caption{A comparison of simulated galaxies with GASS observations 
 from \citet{Catinella-13} for galaxies with $M_*\geq10^{10}\msolar$.
 We show the \HI\ mass (top) and \HI\ richness (bottom) as a function
 of halo mass. The non-detected sources and the corresponding
 simulated galaxies are included in the median. Blue solid lines are running medians
 for simulated galaxies while the red stars show median values for the GASS data.
 Error bars are obtained from cosmic variance (for simulation) and jackknife
 resampling (observations).
 The dashed black lines are running medians for simulated galaxies 
 using a model without quenching (and a slightly different wind model).}
 \label{figsim_obs}
\end{center}
\end{figure}

There are emerging observations of the relationships between \HI\
content in centrals and satellites as a function of halo mass.  In
particular, \citet{Catinella-13} combined data from
GASS$^7$ with the \citet{Yang-07} SDSS
halo catalogue to separate their galaxies into satellites and centrals.
In Figure~\ref{figsim_obs}, we compare $\MHI$ and $\MHI/M_*$ as a function of $M_{\rm halo}$ for the combined sample of
centrals and satellites to these data.
The blue lines show the median values for all the simulated galaxies
(with the error bars showing cosmic variance),
while the red stars show the medians of all the observational data
(with the uncertainties computed using jackknife resampling over 8 subsamples).
To make a fairer comparison to the data, we restrict our
sample to $M_*>10^{10}M_\odot$ galaxies. In their sample,
galaxies with $\log (M_*/\msolar)\leq10.5$ have a fixed \HI~mass lower limit value of 
$\log (\MHI/\msolar) =8.7$ whereas for more massive galaxies, 
the \HI~limit is derived from $\log (\MHI/M_*) = -1.8$; we clip
our simulated values according to those limits and added 
a random scatter of $0.3$, which we adopt to produce a comparable scatter
to that seen in the lower limits within the GASS sample.
In this way, the simulated galaxies having \HI~mass below
the GASS detection limit are still included in the simulated sample, with their
gas content increased to the observational detection limit. However, 
because we are considering medians, this does
not have a major impact on the results, except in the highest mass bin 
where the
median simulated galaxy is a non-detection according to the GASS limit.  In this case, had we taken the
actual $M_{\rm HI}$ value rather than the GASS limit, the medians of 
both $M_{\rm HI}$ and \HI\ richness would be lowered by $\sim 1$ dex.

\begin{figure}
 \begin{center}
 \includegraphics[scale=0.4]{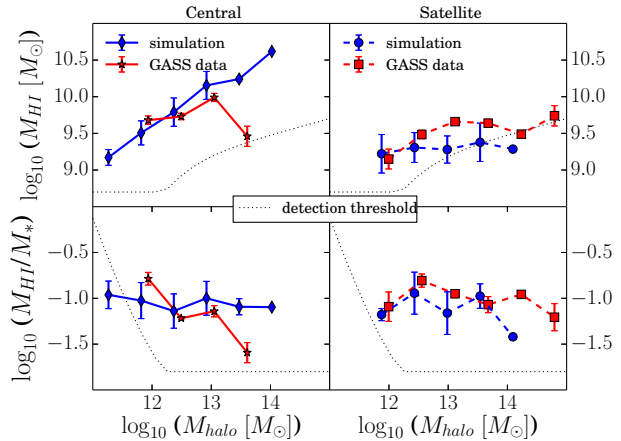}
 \caption{Similar to Figure \ref{figsim_obs} showing \HI\ richness
 versus halo mass, here excluding galaxies below the
 observational detection limit, and also showing the breakdown between central
 and satellite galaxies. Blue lines represent the simulation
 (\textit{diamonds}: centrals; \textit{circles}: satellites) and red lines the
 GASS data (\textit{stars}: centrals; \textit{squares}: satellites). 
 Considering the rather small stellar
 mass range and the limited \HI~mass, all of the medians show a flat
 relationship between \HI-richness and halo mass. The simulations generally
 agree with the observations except for central galaxies within
 $\geq10^{13}\msolar$ halos, though there are modest systematic offsets
 in the satellite populations. The black dotted lines approximately shows the
 GASS detection threshold.}
 \label{figsim_obs_new}
\end{center}
\end{figure}

Figure~\ref{figsim_obs} shows that the models qualitatively
track the data across the halo mass range probed, but
there are systematic discrepancies.  For the observations, the
uncertainty in the median is computed via jackknife resampling over
8 subsamples, which yields an extremely small statistical uncertainty.
These do not include potential systematic errors such as cosmic
variance and uncertainties in the measurements themselves, which
would likely make the errors significantly larger.  The model error
bars show cosmic variance over the eight simulation sub-octants.
There are also systematic uncertainties in computing the \HI\ content
from our models, as our radiative transfer approximation is necessarily
crude and can introduce uncertainties potentially as large as several
tenths of dex.  Hence, the discrepancy in the predicted $\MHI$ of
no more than 0.2~dex, and that for \HI\ richness peaking at 0.3~dex
in the poor group ($\sim 10^{13}\msolar$) regime, represents encouraging
if not perfect agreement.

While there is general agreement, these discrepancies may prove
enlightening.  In the lower panel, we notice that the \HI~richness
predicted in the simulation is somewhat above the observations (
at most $\sim2\times$ \HI~richer) for halo masses around M$_{halo}<\sim 10^{13}\msolar$,
being the same in the lowest halo mass range. The decrease in gas fraction
from the simulation seems to be lower than that in the observational data.
In the high mass halos ( M$_{halo}> 10^{13}\msolar$), the observational
data show a rather constant \HI~richness regardless of the halo mass,
while that of the simulations decreases with
halo mass owing to the growing population of \HI~poor galaxies as explained
previously and also shown in Figure \ref{figcensat}. However, this is
partially driven by the fact that many of the observed galalxies
in this bin are non-detection placed at their upper limit of
$\log (\MHI/M_*) = -1.8$. While we have tried to mimic GASS selection
as closely as possible, it is difficult to assess the robustness of a
comparison driven by non-detections. However, it may suggest
that the predicted satellite \HI~content in this crucial poor group
mass range is somewhat discrepant with the data. At lower and higher
halo masses, the simulation agrees better with the observations.

So far, we have included the undetected sources in the
comparison mimicking the way GASS handled such cases.  For most
bins this had little effect, but at the massive end there were many
non-detections.  To investigate this, we present in Figure \ref{figsim_obs_new}
a comparison between the simulations and the observations, now
excluding sources that were undetected in \HI~and the corresponding galaxies
in the simulations.  We further separate the sample into central
and satellite galaxies.  The blue lines represent the simulations, with
the diamonds representing the central galaxies and the circles the satellite
galaxies. The red lines represent the observational data, with the stars
indicating the centrals and the squares the satellites.
The error bars show $1\sigma$ uncertainties
in the median within each bin, as described for Figure~\ref{figsim_obs}.
We also show as the black dotted lines the GASS detection threshold,
as mentioned previously.  We assume that a given stellar mass corresponds to a halo mass as
determined by \citet[][see their eq.~7]{Yang-07b} from their SDSS halo catalog, namely:
\begin{equation}\label{starhalo}
M_{*,c} = M_0\times{(M_h/M_1)^{\alpha+\beta}\over{(1+M_h/M_1)^\beta}}
\end{equation}
with $[\log M_0, \log M_1, \alpha, \beta] = [10.86, 12.08, 0.22, 1.61]$.
At high halo masses, the GASS data are mostly near the detection threshold.

The simulation and the observations are again in reasonable agreement
(though not within the formal statistical uncertainties), with the exception
of central galaxies in halos greater than $10^{13}\msolar$. This plot
at face value shows weak environmental effects on the satellite
galaxies, as the median satellite has a roughly invariant \HI~mass
and richness.  However, this is only for the {\it detected} population.
As discussed in Figure~\ref{figcensat}, the simulations predict
that the main environmental effect on satellites consists of a
fairly rapid removal of \HI, which means that the galaxies that retain their
\HI~are not substantially different. However,
the {\it fraction} of galaxies with \HI~is a strong function of halo mass,
as seen in Figure \ref{figcensat}.

Returning to Figure \ref{figsim_obs}, we now examine the impact
of our quenching model on these results by comparing the vzw model
without quenching to our standard ezw model. The vzw model (dashed
lines) is slightly lower than our primary model in both gas content and
\HI\ richness.  This is consistent with the results from \citet{Dave-13},
who found that stronger outflows at low masses, as assumed in our ezw model,
yield greater \HI\ masses in galaxies.  In this case, the selection criterion
(corresponding to M$_{\rm halo}\ga 10^{12}\msolar$) emphasises galaxies that are
most susceptible to being quenched.  Despite the quenching in the ezw model,
the \HI\ content in this model is higher than in the vzw model, which likely
stems from the lower mass loading factors present in the vzw model at low 
galaxy masses, which simply gets propagated to higher 
masses\footnote{ A comparison
between the ezw and vzw models as a function of stellar mass was 
shown in \citet{Dave-13}. }.  Hence, our quenching prescription does
not seem to have a major effect on the \HI\ content of high-mass
galaxies.  Coincidentally, the vzw model actually matches these data
somewhat better, though we will show later that it does significantly
worse in comparisons with other environmental measures.

In summary, the halo mass is a strong determinant for the \HI\
content of its galaxies. For all the simulated galaxies
in $M_{\rm halo}\la 10^{12}\msolar$ halos, the \HI\ mass
in centrals rises with halo mass while the \HI\
richness drops, and the \HI\ richness of satellite galaxies is
essentially unaffected by being within another halo.  In more
massive halos, the \HI\ mass of central galaxies continues to rise but more
slowly, and the median \HI\ richness in satellites drops dramatically,
driven by the bimodal appearance of substantial numbers of \HI-poor
satellites.  This trend is likely driven primarily by gas stripping,
as we discuss further below; our ad hoc quenching model has little
impact on this.  The predicted \HI\ content of satellites in massive
halos is in reasonable agreement with observations when matching
the data selection, with some modest discrepancies particularly
at high halo masses.  The
GASS galaxies show a less strong 
trend versus halo mass~\citep{ Catinella-13}, but when we apply similar detection
limits to our simulations, the agreement is reasonable though not perfect.
While these trends in massive halos
have already been seen in observations, they provide a valuable
test for models, and at low masses our simulations make predictions
for the \HI\ content as a function of halo mass and environment
that could be tested against future surveys.

%---------------------------------------------------------------------------------------------------------------------------
\subsection{Timescales for \HI\ loss in satellites}\label{timeHI}
\indent

\begin{figure}
 \begin{center}
 \includegraphics[scale=0.4]{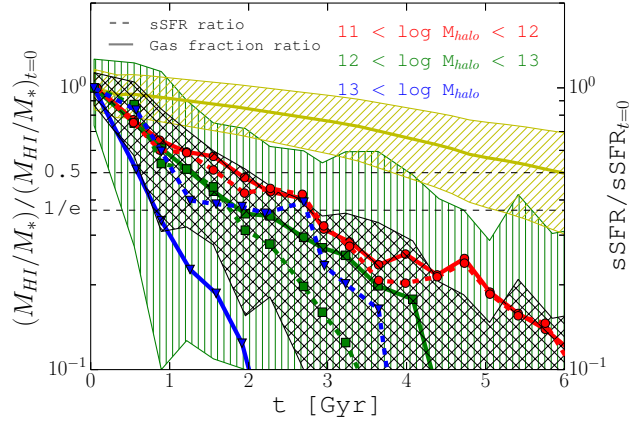}
 \caption{The evolution of \HI\ richness and sSFR for satellite galaxies after infall
 into halos with masses 10$^{11}\msolar\leq$ M$_{halo} <$ 10$^{12}\msolar$
 (red with circles), 
 10$^{12}\msolar\leq$ M$_{halo} <$ 10$^{13}\msolar$
 (green with squares) and M$_{halo} \geq$ 10$^{13}\msolar$
 (blue with downward triangles).
 Thick solid lines show median values for the \HI\ richness relative 
 to the value at $t = 0$, defined for each galaxy as the time when it becomes a satellite (left axis).
 The green vertical hatched area represents the $1\sigma$ dispersion 
 for 10$^{12}\msolar\leq$ M$_{halo}<$ 10$^{13}\msolar$.
 Dashed lines (with the colours and markers corresponding to the halo mass bins)
 correspond to median values for the sSFR relative to the value at t = 0 (right axis).
 The black cross hatched area represents the $1\sigma$ uncertainty of
 the sSFR in host halos of mass $\sim10^{12}\msolar$.
 The yellow line plots the \HI~richness fraction of central galaxies 
 (oblique hatched area $=1\sigma$ scatter) that  have never been
 satellite, with $t=0$ the time when their stellar masses are comparable
 to those of satellites at infall into host halos of mass $\sim10^{12}\msolar$). This is used as a reference.
 Satellites take less time in massive halos to remove their gas.}
 \label{figtime}
\end{center}
\end{figure}

The bimodality of \HI\ richness in satellites of high-mass halos
indicates a fairly rapid loss of \HI\ once a satellite falls into
a $\ga 10^{12}\msolar$ halo.  In this section, we trace
galaxies back in time to determine the evolution of their \HI~content
once they becomes satellite, to quantify the timescales
for \HI\ removal.

In Figure \ref{figtime}, we show the evolution of the {\HI} richness
of satellite galaxies since they were last  central ($t=0$).
We only consider galaxies that were always central before
$t=0$ and that stay satellite until $z=0$. Then, in later outputs,
we compute the \HI~ richness relative to the \HI\ richness at
$t=0$ \footnote{ The scatter at $t=0$ is a plotting artefact owing
to the binning; all values at $t=0$ are unity by definition.}. We plot this
in Figure \ref{figtime} for
three different halo mass bins: 10$^{11}\msolar\leq$ M$_{halo}<$
10$^{12}\msolar$, 10$^{12}\msolar\leq$ M$_{halo}<$ 10$^{13}\msolar$
and 10$^{13}\msolar\leq$ M$_{halo}$, where the halo masses are those
at the time of infall.  The median \HI\ richness as a function of
time is indicated by the red (with circles), green (with squares),
and blue (with triangles) lines for the increasing halo mass bins,
respectively.  The green vertical hatched region represents
the $1\sigma$ variation
around the median for the 10$^{12}\msolar\leq$ M$_{halo}<$
10$^{13}\msolar$ case; the others are similar. The yellow line
(with the $1\sigma$ uncertainty shown by the oblique hatched
area), for reference, shows the \HI~richness of central
galaxies that have never been satellite, where $t=0$ is the time
when the galaxies' stellar masses are comparable to the satellites'
stellar masses upon infall into a $\sim10^{12}\msolar$
halo.

Satellite galaxies in massive halos take less time to have their \HI\
removed than lower halo mass galaxies.  There is only a small
difference between 10$^{11}\msolar\leq$ M$_{halo}<$ 10$^{12}\msolar$
and 10$^{12}\msolar\leq$ M$_{halo}<$ 10$^{13}\msolar$
 halos, having a timescale of 1-2~Gyr
to be lowered to half of their initial \HI\ richness, while for
$\geq10^{13}M_\odot$ halos the timescale is significantly
less than a Gyr.  Hence, gas removal begins quite rapidly, often in
less than a single halo dynamical time, but nonetheless full stripping
of the \HI~(e.g. less than 10\% of initial \HI~left) does not occur
until after several Gyr.  Indeed, in the lowest mass bin, even after 5~Gyr 
the \HI\ is still typically $\ga 10\%$ of the initial value.
Note that the scatter about the median is quite large, so that the
typical time to lose half of the \HI\ at $1\sigma$ above the median
is $> 3$ Gyr instead of $\sim$1.5~Gyr.  Hence, at least some satellites
continue to retain significant \HI\ for many Gyr even in high mass
halos. It is also worth mentioning that there is a significant
difference even between the satellite galaxies in the lowest halo
mass bin and the central galaxies having similar stellar mass (yellow
region), showing that there is some environmental effect on satellites
even in lower-mass halos. We note that some SAMs have adopted the
prescription that hot gas is fully stripped immediately or
within a halo dynamical time when it enters into a massive
halo~\citep[e.g.][]{DeLucia-12}, but this is an oversimplification
that is unlikely to be correct even in the median case.
Recent SAMs have begun to include more sophisticated descriptions
of gas stripping~\citep[e.g.][]{Henriques-15}.  Finally, it is
important to recall from Figure~\ref{figcensat} that the distribution
of \HI\ richnesses in satellites is bimodal, meaning that the median
evolution more broadly tracks the {\it fraction} of satellites that
are having their \HI\ stripped, rather than the evolution of any individual
satellite.  Hence, the timescales here should be viewed as the typical
timescale after infall at which \HI\ is stripped, but the stripping
itself may happen more rapidly.

Since the decay is roughly log-linear, we can fit an exponential
decay timescale for the \HI\ richness of the form $R_{\rm
HI}=e^{-t/\tau}$, where $R_{\rm HI}$ is the ratio of $\MHI/M_*$
relative to that at $t=0$.  For the \HI\ richness, the $e$-folding
decay timescales are $2.63, 2.01, 0.70$~Gyr for the
10$^{11}\msolar\leq$ M$_{halo}<$ 10$^{12}\msolar$,
10$^{12}\msolar\leq$ M$_{halo}<$ 10$^{13}\msolar$ and
10$^{13}\msolar\leq$ M$_{halo}$ bins, respectively, thus quantifying
the more rapid \HI\ loss in massive halos.  In low mass halos,
the timescales are comparable to or longer than the halo dynamical
time of $\sim 2.5(1+z)^{-3/2}$~Gyr \citep{Finlator-Dave-08},
indicating that gradual processes such as starvation are the drivers,
while in more massive halos it is much shorter than a halo dynamical
time, suggesting a more efficient and local process.  When taking
a fix stellar mass range $10^9\msolar\leq M_*\leq10^{10}\msolar$
at $z=0$ (chosen to ensure that there are still many galaxies
at the time of infall above the resolution limit), the e-folding
\HI~richness decay timescales become $2.88, 1.84, 0.26$~Gyr for the
10$^{11}\msolar\leq$ M$_{halo}<$ 10$^{12}\msolar$,
10$^{12}\msolar\leq$ M$_{halo}<$ 10$^{13}\msolar$ and
10$^{13}\msolar\leq$ M$_{halo}$ bins. Particularly for galaxies with
$10^9\msolar\leq M_*\leq10^{10}\msolar$ 
in massive halos (at $z=0$), it takes
a very short time for their \HI\ to be stripped.

To better understand the physical origin of the decline in \HI, we
can compare the decay in \HI\ richness to that of the star formation
rate.  If the decline in \HI\ is much faster than the decline in
SFR, then this indicates preferential removal of \HI\ relative to
star-forming gas, suggesting stripping.  If the decline rates are
comparable, then this can be most simply interpreted as starvation.
In the absence of environmental effects, galaxies are in a steady
state of accretion versus
consumption~\citep[e.g.][]{Dekel-09,Dave-Finlator-Oppenheimer-12,Lilly-13}.
But if the accretion is truncated owing to environmental effects,
leading to starvation, then as the \HI\ runs out, the star-forming
gas will deplete, and hence the SFR will commensurately drop.

In Figure \ref{figtime}, the dashed lines analogously show the
evolution of the specific SFR ($\equiv {\rm SFR}/M_*$) in satellites
once they fall into halos (the colours show the respective
halo mass bins).  For the lower mass halos, the timescale for sSFR
decay is virtually identical to that of \HI\ richness decay (though
we see a small discord between sSFR and \HI-richness for the
$10^{12}-10^{13}\msolar$ halo mass bin after 2~Gyr).
This indicates that
these galaxies are simply consuming their gas and it is not being
replenished, resulting in a coincident drop in gas and SFR.  Hence
in $M_{\rm halo}\la 10^{12}M_\odot$ halos, starvation appears to
be the key mechanism for \HI\ attenuation, which occurs because the
streams that feed low-mass halos generally feed the central galaxy
rather than satellites. This is likely the origin of the residual
environmental effect that is evident from the difference in the 
satellites' and centrals' evolution (red or green vs. yellow line).

% \citet{Wetzel-13} examined the decline in
% satellite SFR in (generally) higher mass halos, and also found that
% the timescale for SFR truncation is long, typically 2$\sim$4 Gyr.
% They also argued for a ``delayed-then-rapid'' decrease in
% sSFR, such that after infall, the sSFR remains mostly unchanged for
% this time after which it drops rapidly within a dynamical time $<<$
% 1Gyr.  This delayed-then-rapid scenario is highly reminiscent of
% what our models predict for the \HI~content, which shows a typical
% decay timescale of several Gyr but then rapidly loses its \HI\
% resulting in a bimodal \HI\ richness distribution (cf. Figure~4).

For high mass halos, in contrast, the \HI\ attenuates much more
quickly than the sSFR.  The post-infall evolution of sSFR is
essentially independent of halo mass\footnote{Figure \ref{figtime}
shows that the evolution of the sSFR ratios are all within $1\sigma$
uncertainty of $10^{12}\msolar\leq M_{halo} < 10^{13}\msolar$.},
indicating that the denser gas and galaxy environment around
more massive halos do not markedly affect gas consumption from an existing
star-forming gas reservoir.  The more rapid drop in \HI\ suggests
that ram pressure stripping is at work, since this should affect
the loosely-bound \HI\ much more than the dense star-forming gas.
Hence the onset of a hot gaseous halo, even in fairly poor groups
with $M_{\rm halo}>10^{13}M_\odot$, is already sufficiently to
produce substantial ram pressure stripping~\citep{Cunnama-14} of
infalling \HI.  Nonetheless, the SFR drop is mostly unaffected by
ram pressure stripping, and remains consistent with simple starvation
even in massive halos.  The difference between the physics that
results in the decline in SFR versus the decline in \HI\ in massive
halos is an important prediction of these simulations, and can
perhaps reconcile conflicting results regarding the ubiquity of ram
pressure stripping in \HI\ cluster studies~\citep[e.g.][]{vanGorkom-03}
versus the much longer decay timescale of satellite star formation
in massive halos~\citep[e.g.][]{Wetzel-13}.

 \begin{figure*}
 \begin{center}
 \includegraphics[scale=0.5]{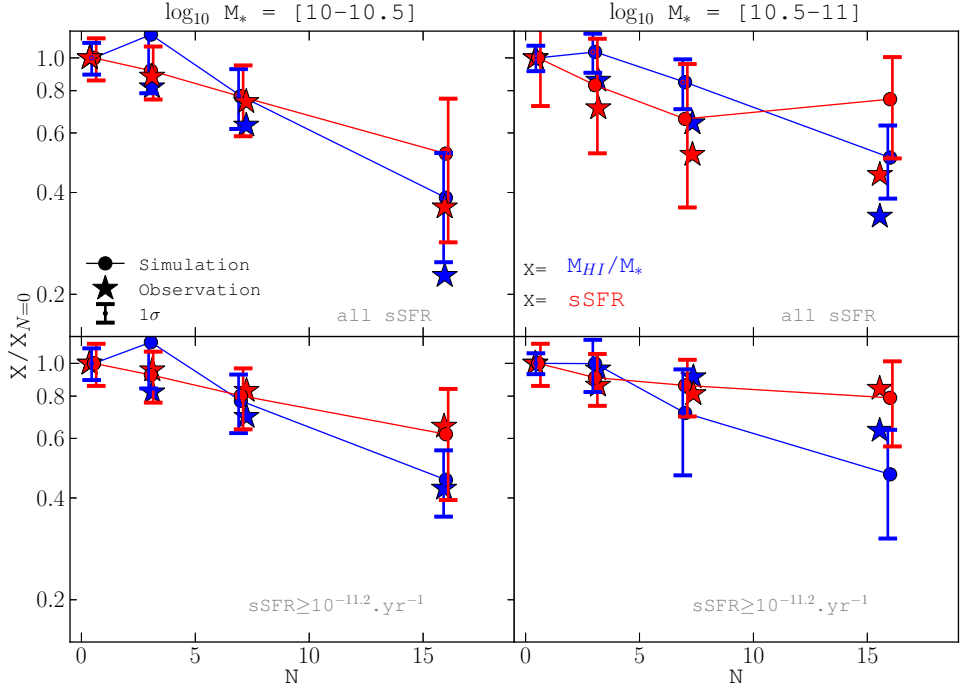}
 \caption{ The median \HI\ richness (blue) and sSFR (red) of galaxies 
 as a function of environmental parameter N (see text) relative 
 to their value for N = 0 galaxies.  Solid lines (with circle markers) correspond
 to our simulated galaxies and  stars to observational data from \citet{Fabello-12}.
 The errorbars are $1\sigma$ uncertainties.
 Left and right panels correspond to galaxies with stellar masses of
 $10^{10}\msolar$\textless M$_*$\textless$10^{10.5}\msolar$
 and $10^{10.5}\msolar$\textless M$_*$\textless$10^{11}\msolar$, respectively.
The bottom panels only include galaxies with sSFR $\geq 10^{-11.2}$ yr$^{-1}$.
 The simulations show a good agreement with the observational data 
 on the \HI~ content fraction for star-forming galaxies but they show 
 some excess in \HI~richness for high N in quenched halos.}
 \label{figngal}
\end{center}
\end{figure*}

%---------------------------------------------------------------------------------------------------------------------------
\subsection{\HI\ as a function of environment}\label{envHI}

In section \ref{csHI}, we found that the \HI\ content of galaxies
is reduced at high halo masses.  Here we examine a related quantity,
which is the \HI\ content as a function of environment.  Since halo
mass is correlated with the environment, e.g. as measured by the number
of nearby galaxies, the reduction in \HI\ at high halo masses is
expected to be qualitatively mimicked in high density regions.  Such
a trend agrees with the observations of \citet{Fabello-12},
which are based on stacked ALFALFA \HI~data. However, they inferred that
environmental processes begin at M$_{halo}\sim10^{13}\msolar$, whereas our
simulations predict that the suppression starts at M$_{halo}\sim$
10$^{12}\msolar$.  In this section we conduct a more detailed
comparison to \citet{Fabello-12}, to better understand
what these data may be telling us about the impact of environment
on \HI\ content.

Figure \ref{figngal} shows the median \HI\ richness and specific SFR
of galaxies as a function of their environment parameter $N$ from
\citet{Fabello-12}, relative to the \HI\ richness and sSFR of $N=0$
galaxies.  $N$ is defined to be the number of galaxies with
$M_*>10^{9.5}M_\odot$ within a cylinder of radius 1~Mpc and a redshift
path length of $\pm 500\kms$. In our simulation, the mean halo mass
for N$\geq$7 (including galaxies with stellar masses 
$10^{10}\msolar\leq M_*\leq10^{10.5}\msolar$) is
$\sim$10$^{13.2}\msolar$, which is consistent with Figure 8 in
\citet{Fabello-12}. We choose two galaxy stellar mass bins,
$M_*=10^{10}-10^{10.5} \msolar$ and $M_*=10^{10.5}-10^{11} \msolar$
to compare to the observational data presented in \citet{Fabello-12}.
In Figure \ref{figngal}, blue represents the \HI\ richness, and red the sSFR,
with the solid lines (with circle markers) showing results from the
simulation and the  stars those from the observations.
The errorbars represent $1\sigma$ uncertainties.  In the upper panels
we apply no minimum sSFR selection criteria, while in the lower panels we select
only the galaxies with sSFR~$\geq10^{-11.2}$yr$^{-1}$, i.e.  it excludes
quenched galaxies. Note that the observational data
determine the global sSFRs from SED fitting over the whole galaxy.
%\footnote{which should only have minor differences versus
%using the SDSS fiber aperture-derived sSFR as in the observations.}.

In both the simulations and data, there is an overall trend of
dropping median \HI\ richness and sSFR in denser environments.
However, the simulation results are typically slightly higher than
the observations particularly at high-$N$, implying that
galaxies located in dense regions have somewhat too much \HI.
This discrepancy is essentially all being driven by the quenched galaxies;
in the lower panels, we see that for star-forming galaxies, the
predicted \HI\ trends of the simulation are in very good agreement
with the data, showing much less decline in dense regions.  

The discrepancy in dense regions may be related to the excess in
\HI\ richness at $M_{\rm halo}\sim 10^{13} M_\odot$ seen in
Figure~\ref{figsim_obs}.  This further indicates that, while overall
the simulation broadly attenuates the \HI\ content as observed, in
detail there may be some discrepancies, especially in our quenched galaxies.

%___________________________________________________
 \begin{figure}
 \begin{center}
 \includegraphics[scale=0.45]{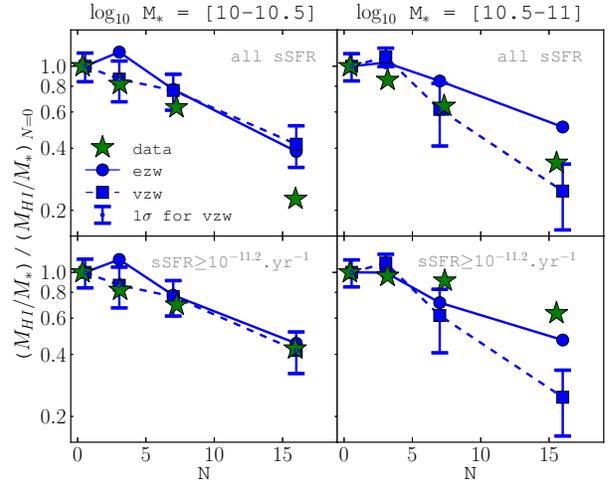}
 \caption{Similar to Figure \ref{figngal}, showing mean \HI\ richness
 as a function of $N$ relative to $N=0$ galaxies, for
 two simulations:
 \textit{Solid lines with circles} correspond to our fiducial model (ezw)
 and \textit{dashed lines with squares} correspond to a model
 without quenching (vzw).  We also show observational data
 from \citet{Fabello-12} as represented by the \textit{stars}.
 The effect of quenching is qualitatively stronger at high 
 $N$ than at high halo mass (cf. Fig.~\ref{figsim_obs})}
 \label{figngalezwvzw}
\end{center}
\end{figure}

A comparison with our vzw model, which excludes quenching, in
Figure \ref{figngalezwvzw}, shows the effect of quenching
for a range of environmental conditions.
Without quenching (vzw), the change in \HI\ richness relative to
the \HI\ richness at $N=0$  towards denser regions is more pronounced
(for $10.5<\log_{10}(M_*/\msolar)<11$ ).  This shows that the quenching
model has a comparable qualitative impact with respect to environment
as it does with respect to halo mass (cf. Fig.~\ref{figsim_obs}).
For lower mass galaxies ($10<\log_{10}(M_*/\msolar)<10.5$), there is
very little difference between the quenching (ezw) and no-quenching
(vzw) simulations, since the quenching mass scale we impose is at the upper
end of this mass range.  For more massive galaxies (right panels),
however, the difference between the two is larger.  The vzw
model actually fares better here, suggesting again that the quenching
model in ezw is not in good agreement with the \HI\ observations.  The ezw model
does fare better, however, than the vzw model for the star-forming sample,
which is not surprising as it produces a better match to the stellar mass
function below $L^*$ \citep{Dave-13}. Given the ad hoc nature of our quenching
prescription, one should not read too much into these trends, but
an important broader point is that observations of \HI\ versus
environment could potentially be used to constrain models of star
formation quenching in central galaxies.

%---------------------------------------------------------------------------------------------------------------------------
\subsection{{\HI} radial halo profiles}\label{rhaloHI}
\indent

So far we have focused on the properties of the total \HI\ content
within galaxies and halos.  To investigate the physical processes
by which environment affects \HI\ in more detail, here we study the
radial distribution of \HI\ within halos.

\begin{figure}
 \begin{center}
 \includegraphics[scale=0.4]{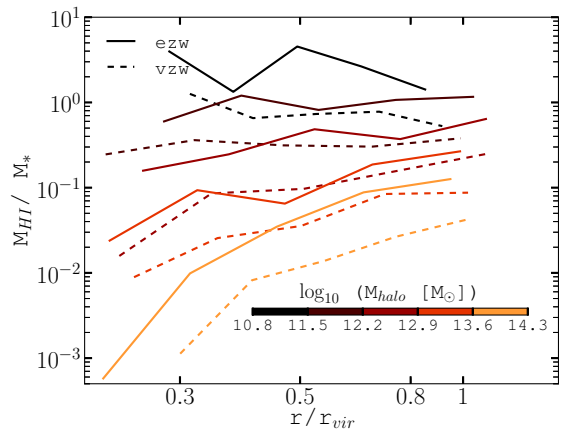}
 \caption{The \HI\ richness of satellites as a function of radial distance
 from the halo centre in units of the virial radius.
 Different colours show mean values for different halo mass bins,
 as indicated by the colourbar. The solid lines show our fiducial model (ezw)
 and the dashed lines show the model without quenching (vzw).
 The \HI~richness of satellite galaxies
 is independent of radius for smaller halos, but 
 decreases by $\sim 100$ times near the centre of a massive halo
 ($M_{\rm halo}\sim10^{14}\msolar$) compared to the outskirts.}
 \label{fighisr}
\end{center}
\end{figure}

Figure \ref{fighisr} shows the \HI\ richness of satellites as a
function of the radial distance from the halo centre, in units of
the halo's virial radius, defined in \S\ref{simu1}
(similar to what was done in \citealt{Solanes-01}). 
The main trend shows that at all radii,
the \HI\ richness always decreases with increasing halo mass.  This
is the same overall trend as shown in Figure \ref{figcensat}, except here we 
additionally show that it occurs as a function of radius.

Besides this overall reduction, there is a noticeable change in the
\HI\ richness profile with halo mass.  In the low mass halos ($M_{\rm halo}\la 10^{12}M_\odot$), we see that the \HI\ richness of satellite
galaxies is independent of their distance to the centre of the halo,
consistent with there being essentially no stripping of their gas
content as the satellite falls deep into these halos.  In contrast,
in high mass halos there is a clear decrease in \HI\ richness at
small radii, with the trend gradually changing from no radial trend
to a very steep one with halo mass.  For the most massive halos
($M_{\rm halo}\sim 10^{14}M_\odot$), satellite galaxies close to
their centre are more than two orders of magnitude lower in \HI\
richness than those near the virial radius, with the trend mildly
steepening within the inner third of the virial radius.  This
decrease of \HI\ fraction for the most massive halos may owe to the
fact that galaxies closer to the centre of the halo accreted earlier
than those at the outskirts, leading to a longer time for
stripping to take effect.

We also show the results from vzw model in Figure \ref{fighisr}
as dashed lines, to check the impact of quenching on the \HI-richness
profile. At all radii and for most of the halo mass range, vzw-galaxies
have approximately half an order of magnitude lower \HI-richness
than ezw galaxies.  Hence the radial trend does not seem to be much
affected by our quenching prescription.  This is not surprising because
our quenching prescription does not directly affect satellite galaxies
since they are typically below our $\sigma$ quenching threshold.

In summary, the radial profile of \HI\ richness shows no trend for
low-mass halos, but at $M_{\rm halo}\ga 10^{13}M_\odot$ there is
a marked drop in \HI\ richness towards the centre which becomes
rapidly more pronounced at higher masses.  This is qualitatively
consistent with the idea that gas stripping is quite efficient in
high mass halos that are abundant in hot hydrostatic gas.  Further
comparisons to observations will elucidate whether these simulations
are capturing all the relevant processes accurately.

%---------------------------------------------------------------------------------------------------------------------------
\section{\HI\ evolution and mergers}\label{evmerg}

%---------------------------------------------------------------------------------------------------------------------------
\subsection{Evolutionary tracks}\label{evtrack}

\begin{figure*}
 \begin{center}
 \includegraphics[scale=.9]{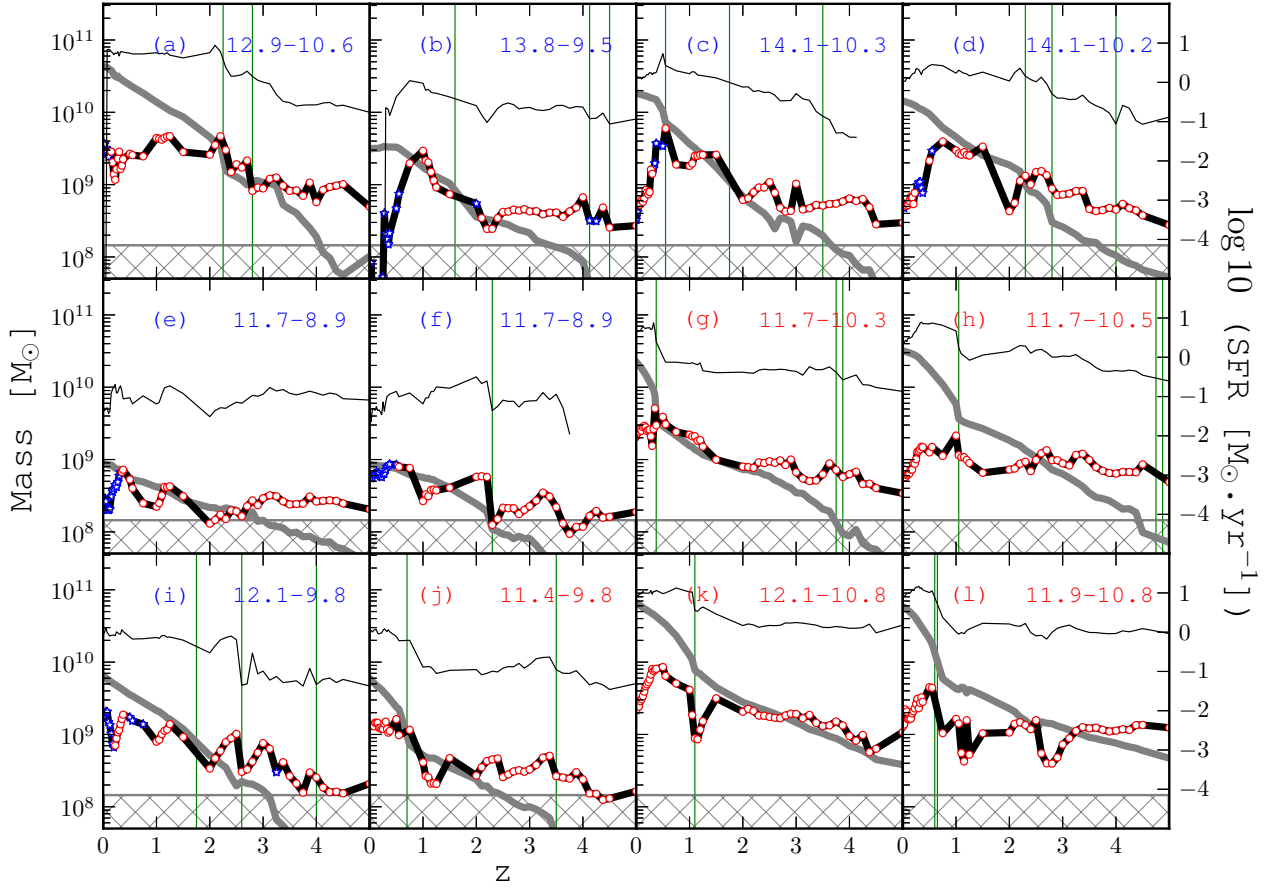}
 \caption{The growth of 12 individual galaxies from z=5 to z=0.
 The thick black lines show the \HI\ mass evolution, the thick grey lines show the stellar mass evolution,
 and the thin black lines show the SFR (scale on the right axis).
 The green vertical lines indicate the redshifts at which galaxies have undergone
 a $>1:3$ merger. The numbers on the top right corner are the halo mass
 and stellar mass at z=0, respectively (the colour indicates whether the galaxy
 is central (red) or satellite (blue) at z=0). Red circles along the \HI~mass
 lines represent that the galaxy is a central at the given redshift and blue stars
  show that the galaxy is a satellite.}
 \label{fighistory}
\end{center}
\end{figure*}

Figure~\ref{fighistory} shows evolutionary tracks from $z=5\rightarrow
0$ for selected individual simulated galaxies spanning a range of
final stellar masses and halo masses.  The galaxies are selected
to show different features such as different final stellar masses,
different halo masses, the effects of quenching and recent and early
mergers.  The thick black lines show the \HI\ mass evolution,
the thick grey lines show the stellar mass growth and the
thin black lines show the SFR evolution (scale on the right axis).
We indicate along the \HI\ mass lines (thick black lines) whether
the galaxy was central (red circles) or satellite (blue stars)
at that epoch.  Vertical green lines show where greater than $1:3$
mergers occur, as described in \S\ref{simu3}. We also show the log
of the halo and stellar masses at $z=0$ by the two numbers in the
top right corner, where the colour of the numbers indicates whether
the galaxy is central~(red) or satellite~(blue) at that redshift.
Note that there are two roughly Milky Way-sized galaxies in terms
of halo and stellar mass, in panels (k) and (l).  

The tracks show that galaxies grow in \HI\ and stellar mass together,
although in general the stellar mass grows more quickly than the
\HI\ mass.  We will show in the next section that in these simulations,
this evolution is a result of the roughly constant relationship
between \HI\ and stellar mass, and is not driven by redshift evolution
of this relation. That is, as the galaxy grows in $M_*$, it
typically has lower $M_{\rm HI}/M_*$, and thus its \HI\ mass has
grown more slowly than stellar mass.  There is also a relationship
between \HI\ growth and SFR, and in particular a merger (identified
by the vertical lines) sometimes drives a rise in the SFR but can
result in either a rise or decline in \HI\ mass; we will quantify
this in \S\ref{merg}.  Another aspect depicted is the quenching
prescription, as described in \S \ref{simu1}, which stops galaxy
from forming stars when it is located in a massive halo. For instance,
in panel (a), we can see that the galaxy is quenched just before
$z=0$ once it entered a massive halo, but its \HI\ gas is still
available to possibly feed star formation before complete
stripping occurs.  Conversely, in panel (b), the SFR starts
decreasing in concert with the \HI, and then eventually is fully
truncated as the \HI\ is removed.  The galaxy ceases to form stars
and the stellar mass even decreases as some of the stars
get stripped. Stripping is also seen in panels (c) \&
(d). In these two cases, the SFR does not go to zero, but we see that
once the galaxy crosses into a bigger halo the gas is quickly
exhausted while the SFR decreases at a far lower rate: this confirms
what we showed in Figure~\ref{figtime} with the blue solid and the
blue dashed lines for the most massive halos.

For central galaxies becoming satellites by entering into
low-mass ($\la 10^{12}M_\odot$) halos, Figure~\ref{figtime} showed
that the decrease in gas fraction is very much similar to the
decrease of specific SFR. To illustrate this, two central galaxies
that have become satellites in less than $10^{12}\msolar$ host halos
are shown in panels (e) \& (f): the
\HI\ content decreases along with the SFR.  We speculate that these
galaxies do not receive any gas infall, and hence the decrease in
their gas content owes to the fact that they are still producing
stars without much further gas replenishment.  Galaxies like
(i) show highly fluctuating gas content as well as star formation
rate,  owing to mergers or fly-by's where some of the particles of
either of the galaxies get disrupted or even stripped from the
galaxies. The two Milky Way-like galaxies (k) and (l) both undergo
mergers at fairly late epochs, but in (l) the post-merger
galaxy then reduces its \HI\ mass, whereas in (k) it gradually
increases after the merger. The galaxy in panel (j) undergoes
a merger at $z=0.7$ but, at least down to $z=0$, the \HI\ richness
seems unaffected by this.

These tracks illustrate the diversity in trends that impact the
\HI\ evolution in galaxies that we explore next, and in particular 
the relationship between mergers and \HI\ that we quantify below.

%---------------------------------------------------------------------------------------------------------------------------
\subsection{\HI\ richness evolution with redshift}\label{evHI}

Our results above suggest that massive halos with hot gas are
effective at stripping the \HI\ from infalling satellites, while
$M_{\rm halo}\la 10^{12}M_\odot$ starve their satellites of \HI\ over longer
timescales.  This explains the trend in \S\ref{csHI} where the \HI\
richness drops significantly for satellites relative to centrals
in halos above this mass.  In this section, we examine at which
cosmic epoch this trend appears, and how it evolves to the present
day.

\begin{figure}
 \begin{center}
 \includegraphics[scale=0.4]{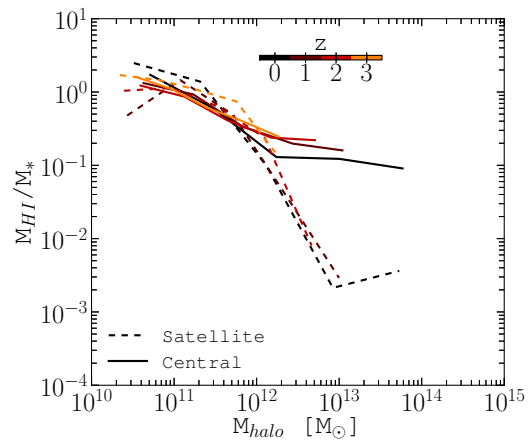}
 \caption{The median \HI\ richness as a function of halo mass at different redshifts,
 from z=3 (orange) to z=0 (black).
 Solid and dashed lines correspond to central and satellite galaxies, respectively.
 Different populations of galaxies at different redshifts show no redshift
 evolution of \HI~richness with respect to host halo mass.}
 \label{fighirhalo}
\end{center}
\end{figure}

Figure \ref{fighirhalo} shows the evolution of the median \HI\
richness versus halo mass (only including galaxies with $\resgal$
at the respective redshift), similar to that
in the lower panel of Figure~\ref{figcensat} but now showing the redshift
evolution from $z=3$ to $z=0$.  In this plot, the stellar mass
and halo mass of the galaxy are measured at the respective redshift.
We take all of the galaxies regardless of whether they are progenitors
of those at z=0 or not as long as $\resgal$.
As discussed in \citet{Dave-13}, the \HI\ richness at a given mass 
hardly evolves in time; in \citet{Dave-13} this was shown versus 
stellar mass, while here we show it versus halo mass, 
but as expected from abundance matching models
\citep[e.g.][]{Behroozi-Wechsler-Conroy-13} there is not
much evolution in the $M_*-M_{\rm halo}$ relation.  At all redshifts,
we see the trend of lower \HI-richness in more massive systems, which
continues roughly unabated over the mass range probed.

In contrast to centrals, satellites at all redshifts show a reduction
in their \HI\ richness relative to centrals at $M_{\rm
halo}>10^{12}M_\odot$.  This is likely tied to the emergence of hot
gaseous halos at this mass scale, which is roughly independent of
redshift~\citep{Keres-05}.  At the highest redshifts probed here
($z=3$), we do not have sufficient volume to probe satellites in
halos significantly above this mass, but it seems that the trend
remains consistent.  We note that \citet{Dave-13} demonstrated that
the \HI-poor fraction of satellites occurs regardless of the
quenching prescription (which is based on $\sigma$, and so
lower-$\sigma$ satellites are generally unaffected); hence, this is
not what is driving the decrease.  Rather, the processes described
in the previous section of starvation and ram pressure stripping
are predominantly responsible.

\begin{figure}
 \begin{center}
 \includegraphics[scale=0.4]{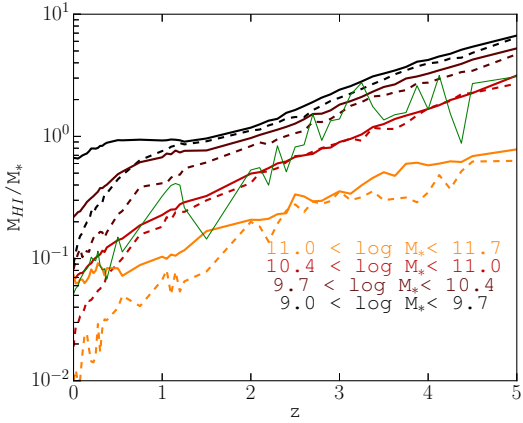}
 \caption{The median \HI\ richness as a function of redshift
 for central galaxies (solid lines) and satellites (dashed lines)
 in different stellar mass bins defined at $z=0$, increasing from
 M$_*\sim$10$^{9.3}\msolar$ (black) to M$_*\sim$10$^{11.3}\msolar$ (orange).
 The green line shows the median \HI~richness of all galaxies
  that are still satellites at that redshift, with $\resgal$.
 }
 \label{figrunmed}
\end{center}
\end{figure}

Figure \ref{figrunmed} shows the median {\HI} richness of central
and satellite galaxies from $z=5-0$.  We separate the galaxies into
bins of $z=0$ stellar mass as shown in the figure by the
colour coded lines, and compute the medians for all those galaxies'
main progenitors at each redshift.  Hence, each line defines a fixed
population of galaxies, although their environment varies over time.
Here we expand our simulated sample at high-$z$ to masses
below our nominal resolution limit to extend the curves
to high redshifts. If we were to limit our progenitors to only
resolved galaxies the curves would not show a different trend, but
would only be noisier.

The predominant trend is that galaxies are more \HI-rich at early
times.  However, at $z\ga 1.5$, the difference between
satellites and centrals is very small, and the two track each other
very well.  This could be because many of those satellites
were centrals in their own halo at early times.  To check this, we
also show as the green line the median \HI~richness for only the
galaxies that are satellites at each redshift, and that are above
our $M_*$ resolution limit.  Note that this is not a fixed population,
as there are fewer galaxies that are still satellites at higher
redshifts, and hence the scatter gets larger at higher redshifts.
Nonetheless, the trend of increasing \HI~richness is generally
present at roughly the same rate for galaxies that are satellites
at every epoch.  This indicates that even though some satellites
may be centrals at earlier epochs, in fact the satellites themselves
are also more \HI~rich at higher redshifts.

At $z\la 1.5$, the satellite galaxies' \HI\ richness begins to
depart significantly from that of the centrals in halos of all
masses, as large potential wells grow that can more effectively
strip satellites.  Observational \citep{Muzzin-14} and theoretical
\citep[e.g.][]{Bahe-15} studies also suggest that gas stripping can
be effective in cluster mass halos starting at these early epochs.
An interesting feature we see is the flattening in \HI\ richness
evolution at $z\la 1$ for the lowest mass central galaxies; the
origin of this is not entirely clear.  The lowest mass satellites,
meanwhile, show the strongest drop to $z=0$, as they are most
susceptible to stripping processes.

 \begin{figure}
 \begin{center}
 \includegraphics[scale=0.4]{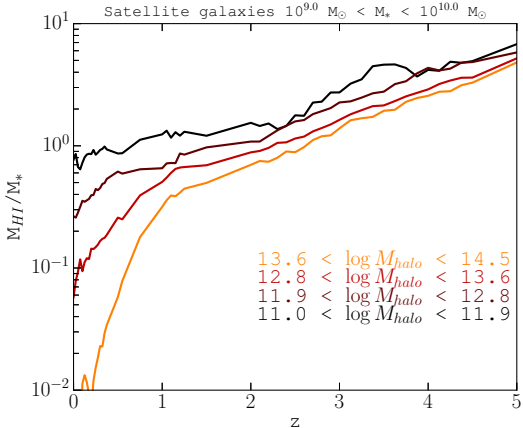}
 \caption{The median \HI\ richness as a function of redshift 
 for satellites in different $z=0$ halo mass bins,
 increasing from M$_{\mbox \tiny halo} \sim 10^{11.5}\msolar$ (black)
 to M$_{\mbox \tiny halo} \sim 10^{14}\msolar$ (orange).
 We only include satellites in the stellar mass range
 10$^{9}\msolar$\textless M$_{*}$\textless10$^{10}\msolar$ at z=0
 and their most massive progenitors at higher redshifts.}
 \label{figrunmedhalo}
\end{center}
\end{figure}

We can also examine the evolution in bins of halo mass rather than
stellar mass, as shown in Figure~\ref{figrunmedhalo}.  We focus
here on low-mass satellites since they are most strongly impacted
by \HI\ removal processes, namely satellites with
$10^{9}\msolar<M_*<10^{10}\msolar$.  We choose this mass range
because, while we resolve to lower masses at $z=0$, we cannot track
their history very far back in time before they fall below our
resolution limit.  Figure~\ref{figrunmedhalo} plots the {\HI}
richness evolution from $z=5\rightarrow0$ for these satellites in
various $z=0$ halo mass bins as indicated by the colour-coding.
Again, we see that at $z\la1$ there is a marked drop in the \HI\
richness of satellites, but only in the most massive halos ($M_{halo}\ga
10^{13}M_\odot$).  The redshift at which \HI\ removal becomes
important is a strong function of halo mass, consistent with the
idea that stripping begins when the halo mass crosses a mass threshold
(Figure~\ref{fighirhalo}) coincident with the onset of a substantial
hot gaseous halo.

%---------------------------------------------------------------------------------------------------------------------------
\subsection{\HI\ in mergers}\label{merg}
\indent

In this section we investigate the impact that major mergers have
on the \HI\ content of galaxies, and in particular we quantify how
much scatter in the relationship between $\MHI$ (or \HI\ richness)
and $M_*$ is induced by mergers.

Figure~\ref{fighistory} shows that mergers in many 
cases result in a burst of star formation.  What this means for the
\HI\ is unclear: we might expect a decrease of \HI\ content owing
to gas consumption that is too rapid for replenishment, but it could
be that the \HI\ might be increased if this gas is fuelling the
burst. For instance,
in Figure~\ref{fighistory} (c)(at z$\sim$0.5), (f) or (h) we can see
variations in the decrease in gas content. Conversely to those
cases, we also encounter situations where the gas content keeps
increasing, such as in Figure~\ref{fighistory}(c)(at z$\sim$1.75),
(i)(at z$\sim$1.75) or (k).  One possible impact for mergers may,
therefore, be to increase the scatter in the relationship between
\HI\ and stellar mass.  Here we more quantitatively consider 
how mergers alter the \HI\ content of the galaxies.

 \begin{figure}
 \begin{center}
 \includegraphics[scale=0.4]{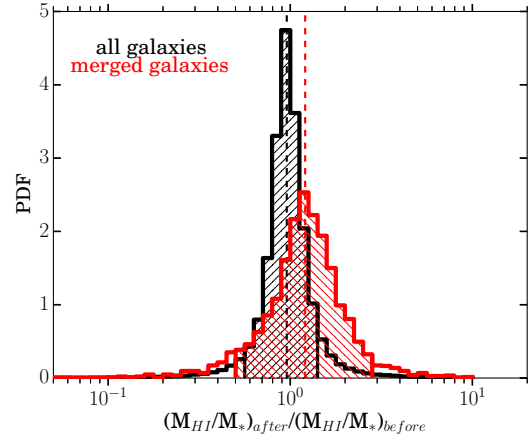}
 \caption{The effects of major mergers on the \HI\ content of galaxies.
 We compare the distribution of the ratio of \HI\ richness after
 and before every output for all galaxies (black histogram) to the distribution
 of ratios after and before every $>1:3$ merger (red histogram).
 Shaded regions contain 90\% of the data for the control (forward slashed)
 and merger (backward slashed) samples,
 with spreads of $\sim0.44$ dex and $\sim0.75$ dex, respectively.
 Median values are indicated by the vertical dashed lines.
 Mergers increase the median \HI~richness of galaxies,
 but the most dominant feature is the increase in the width
 of the distribution interpreted as simple stochasticity.}
 \label{fighist}
\end{center}
\end{figure}

Figure \ref{fighist} quantifies the impact of mergers on the scatter
in \HI\ richness.  Here we show the distribution of the ratios of
the {\HI} richness after and before every output (black histogram)
compared to those after and before a greater than $1:3$ merger (red histogram).
The distribution for all galaxies shows a tiny shift of the
histogram towards values below 1 (median~$\approx 0.95$), which
means that, on average, galaxies have their {\HI} richness decreased
slightly with time. This was anticipated if we refer to
Figure~\ref{figrunmed} and Figure~\ref{figrunmedhalo} where the
\HI\ richness decreases towards low redshift and serves as a
reference for the distribution of the {\HI} richness ratio after and
before mergers.

The post-merger ratio (red histogram) shows two effects.  First,
the \HI\ richness increases, with a median ratio of $1.21$.  This
suggests that mergers generally boost the \HI\ fraction, which
occurs concurrently with an increase in SFR and a decrease in
metallicity~\citep{Dave-13}.  This is consistent with the trend
that high-\HI\ galaxies have recently received a substantial input
of new gas~\citep{Moran-12}, and that such an input also fuels new
star formation.  It is less consistent with the idea that \HI\
gets ``used up" in a merger, even though this situation can be seen
in some cases.  The overall trend of higher \HI\
content in higher SFR galaxies is consistent with
observations~\citep{Robertson-13}.

A second trend from Figure~\ref{fighist} is that the scatter in
relative \HI\ richness is typically larger in post-merger galaxies than in
the overall sample.  This suggests that mergers do indeed tend to increase
the scatter in \HI\ richness at a given mass.  The range of
ratios enclosing 90\% of galaxies rises from $\sim0.44$~dex in
the overall case to $\sim0.75$~dex in the post-merger case.

A useful plot for understanding how \HI\ participates in the baryon
cycle is via a {\it deviation plot}, which quantifies second parameter
variations in relationships versus a single quantity, such as stellar
mass.  An example of such a second parameter trend is the so-called
fundamental metallicity relation~\citep{Mannucci-Cresci-12,
Lara-Lopez-10}, in which galaxies at a given stellar mass show lower
metallicity at higher star formation rates.

\begin{figure}
 \begin{center}
 \includegraphics[scale=0.6]{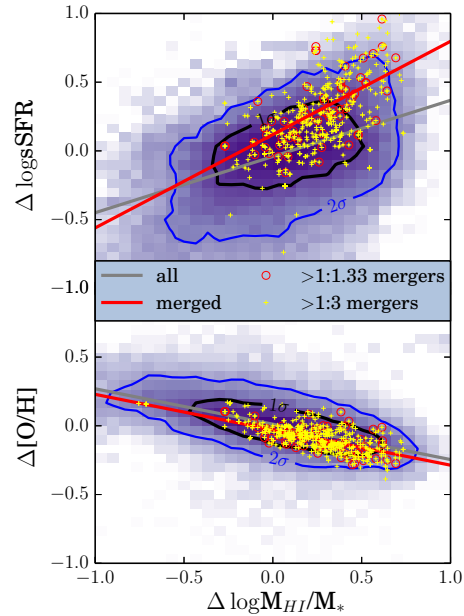}
 \caption{The deviations of sSFR (top) and metallicity (bottom) from
 their median values at a given $M_*$ against the deviation
 of the \HI\ richness from its median value at a given $M_*$ (see text for details).
 The density of galaxies in each deviation plot is indicated
 by the background colour scale, with the black and blue contours containing
 $\sim68\%$ and $\sim95\%$ of the galaxies, respectively.
 Yellow crosses and red circles show deviations for galaxy
 mergers with mass ratios $>1:3$ and $>1:1.33$, respectively.
 The grey and red solid lines show linear fits for all galaxies
 and major mergers, respectively. Merged galaxies sit just above
 the fitted line for all galaxies for the deviation in sSFR.}
 \label{figdeltaplot}
\end{center}
\end{figure}

Figure \ref{figdeltaplot} shows two deviation plots. This quantifies
the deviation of each galaxy from the overall trend given by the
median value in each stellar mass bin.  We first fit a cubic (third-order) spline to the 
median relationship between
SFR and $M_*$, and $M_{HI}$ and $M_*$.  For each value of
$M_*$, we then subtract the quantity from
its corresponding median spline-fit value.  This gives the deviation
for each quantity, which we can then plot against each other.

The upper panel shows the deviation of the sSFR from its median
value at a given $M_*$ ($\Delta\log$ sSFR) against the deviation of
the \HI\ richness from its median value at a given $M_*$ ($\Delta\log
(M_{HI}/M_*)$). The lower panel similarly shows a deviation plot of
metallicity versus \HI\ richness.  To avoid congested plots, we
chose to make density plots with two contours containing 1$\sigma$
(black) and 2$\sigma$ (blue) of the resolved galaxies.  We fit a
power law for all the galaxies (grey line), and for those which
have just undergone a major merger between the last two outputs
(red line).

From the top panel, we can see that recent mergers generally follow a different
relation than the overall galaxy population.  In particular, galaxies
with a high sSFR also tend to have a high \HI\ richness, but after
a merger the sSFR is enhanced {\it more} than the \HI\ richness.
The grey line is similar to what was shown in \citet{Dave-13}, which
is explained that recent gas accretion tends to boost SFR.  The
point of Figure~\ref{figdeltaplot} is that mergers boost the SFR
above the overall relation owing to general stochastic fluctuations
in accretion, more so than the \HI.

For comparison, the bottom panel shows a deviation plot of \HI\
richness versus metallicity.  This shows that galaxies with increased
\HI\ richness typically show a decreased metallicity at a given stellar mass.
In contrast to the sSFR, here the trend for all galaxies is
identical to recently merged galaxies, showing that low metallicity
goes hand-in-hand with higher \HI\ content independent of whether
a merger happened or not.  This is consistent with the idea that the recent
infall that boosts the \HI\ content also brings in lower metallicity gas.
This shows
that \HI\ traces metallicity in the baryon cycle, while the SFR
is specifically boosted by interactions.

Overall, mergers have a minor but noticeable impact on the \HI\
content of galaxies, increasing it immediately after the merger,
and adding to the scatter in \HI\ richness.  However, the increase
in \HI\ is essentially consistent with simple stochasticity in
accretion within the baryon cycle, without there being an {\it
additional} increase owing to the merger as seen in the case of the
SFR.  We caution that owing to the limited time resolution of our
outputs, and possibly the low spatial resolution that is unable
to fully resolve the galaxies' internal structure, this analysis may be
underestimating the impact of mergers.  More detailed simulations,
and associated comparisons with data, can help us disentangle 
the roles of all the components of galaxies in mergers.

%---------------------------------------------------------------------------------------------------------------------------

%---------------------------------------------------------------------------------------------------------------------------

%---------------------------------------------------------------------------------------------------------------------------
\section{Summary}\label{sum}

We investigate the properties of \HI\ in galaxies drawn from a
cosmological hydrodynamic simulation including galactic outflows
that reproduces many key observed properties of the \HI.  We focus on
studying the relationship between \HI\ and environment, quantified
either via halo mass, central versus satellite galaxies, or local galaxy
density.  We also examine evolutionary trends in \HI\ content versus redshift,
and study how mergers influence the \HI\ content.  Our main results are
summarised as follows:

\begin{itemize}

\item Our simulation shows very good agreement with the observed
ALFALFA galaxy \HI~mass function (HIMF) and, furthermore, shows good agreement
with the observed HIMF broken into stellar mass bins based on the GASS data.
This suggests that the scatter between
$\MHI$ and $M_*$ is generally well-reproduced in our simulation, 
providing a second-order test of our models.

\item The \HI\ content of central galaxies is governed primarily
by the halo mass, with a positive correlation that is steeper and
tighter for M$_{halo}$\textless10$^{12}\msolar$.  The median \HI\
richness of centrals decreases slowly with halo mass, which continues
up to the highest masses probed.

\item For satellites, the medians of \HI~and \HI\ richness
show significantly lower values above $M_{\rm halo}\sim10^{13}\msolar$,
with the  emergence of a bimodal distribution in which the drop in the
median is driven by an increasing fraction of satellites essentially
devoid of \HI.

\item When a galaxy falls into a more massive halo, its \HI\ content is
reduced.  This trend occurs at all halo masses.  However, the
trend is strongly accelerated in halos with $M_{\rm halo}>10^{12}\msolar$.
The median $e$-folding timescale for removal is $\sim 2$~Gyr in
lower mass halos, but only 0.7~Gyr in $M_{\rm halo}\approx
10^{13}\msolar$ halos.

\item A comparison to star formation rate attenuation shows that
the SFR is also attenuated once a galaxy becomes a satellite, but
that there is no strong acceleration of this in massive halos.  The
physical implication is that star formation and \HI\ attenuation
are consistent with starvation in lower mass halos, while in more
massive halos, the SFR still drops owing primarily to starvation
while the \HI\ is strongly affected by additional gas stripping
processes associated with the presence of a hot gaseous halo.

\item The \HI\ richness at a given halo mass does not evolve much
with redshift out to $z\sim 3$. There continues to be a rapid drop
in a satellites' \HI\ richness at M$_{halo}>10^{12}\msolar$.
The median \HI\ richness in all galaxies drops from $z=5\rightarrow 0$,
with the satellites tightly tracking the centrals down to $z\sim 1$
and then dropping rapidly below that.  The drop is tightly correlated
with halo mass, starting first in the highest mass halos at $z\ga 1$, 
while low-mass halos show no strong drop in the satellite \HI\ richness
at late times.

\item In low mass halos (M$_{halo}\la 10^{12}\msolar$), the
{\HI} richness of satellite galaxies is independent of their distance
to the central galaxy; whereas at high halo mass
(M$_{halo}\geq10^{13}\msolar$), galaxies become {\HI} poorer towards
the centre, indicative of strong gas stripping processes.

\item Mergers typically cause a modest increase in the \HI\ richness, while
also increasing the scatter in \HI\ richness.  These deviations are
consistent with mergers being an extreme example of a stochastic fluctuation
in accretion, rather than being driven by internal processes
particularly associated with the merger, as is the case with the
SFR.

\end{itemize}

These results show that environment has a major impact on the \HI\
content of galaxies.  However, this impact is mostly confined to
galaxies with M$_{\rm halo}>10^{12}\msolar$.  The
fact that hot gaseous halos tend to develop around this mass scale
in our simulations~\citep{Gabor-Dave-12}, as expected from virial
shock cooling timescale arguments~\citep{Birnboim-Dekel-03}, suggests 
that the presence of hot gas is the main driver for environmental effects,
particularly for satellite galaxies.  The most straightforward interpretation
of this is that ram pressure stripping effects are driving much 
of the \HI\ satellite evolution in massive halos.  But the timescale
and manner in which the gas gets stripped, and its relationship to
the molecular gas and stars, remains less clear.
In future work we will examine
gas stripping processes in more detail, using more physically-motivated
models for star formation quenching and with simulations that use
improved numerical techniques to better model the interaction of 
\HI\ and hot ambient gas.
This work sets the baseline for such future studies,
and highlights a new set of observational comparisons that can
provide crucial constraints on the evolution of gas in galaxies.

%---------------------------------------------------------------------------------------------------------------------------

 \section*{Acknowledgements}
 The authors thank the referee for providing judicious
 comments and suggestions for additional
 work that helped improve this paper.
The authors thank B. Catinella for providing us with GASS data for
comparison and for helpful comments on the draft, and more broadly
the GASS survey team for spurring this work with useful discussions.
M. Rafieferantsoa acknowledges financial support from the South
African Square Kilometer Array.  This work was supported by the
South African Research Chairs Initiative and the South African
National Research Foundation.  This work was supported by the
National Science Foundation under grant number AST-0847667, and
NASA grant NNX12AH86G.  Computing resources were partially obtained
through a grant from the Ahmanson Foundation.

\bibliographystyle{mn2e}
\bibliography{HImika_c}

\end{document}